\documentclass[10pt,conference]{IEEEtran}
\usepackage[utf8]{inputenc}

\usepackage{xcolor}
\usepackage{colortbl}
\usepackage{multirow}
\usepackage{booktabs}
\usepackage[inline]{enumitem}
\usepackage{graphicx}
\usepackage{hyperref}
\colorlet{tableheadcolor}{gray!25} % Table header colour = 25% gray
\colorlet{tablerowcolor}{gray!10} % Table row separator colour = 10% gray

\setlist[enumerate,1]{label=(\roman*), ref=(\roman*)}

\usepackage{caption}
\usepackage{subcaption}
\usepackage{wrapfig}
\usepackage{multirow}
\usepackage{multicol}
\usepackage[T1]{fontenc}
\usepackage{graphicx}
\usepackage{breakcites}
\usepackage{url}
\usepackage{color}

\usepackage{booktabs}   %% For formal tables:
\usepackage{subcaption} %% For complex figures with subfigures/subcaptions
\usepackage{listings}
\usepackage{relsize}
\usepackage{enumerate}
\usepackage{float}
\usepackage{nicefrac}
\usepackage{multirow}
\usepackage{microtype}
\usepackage{wrapfig}

\usepackage{amsmath}
\usepackage{amssymb}
\usepackage{mathtools}
\usepackage{xcolor}
\usepackage{colortbl}
\usepackage[export]{adjustbox}
\usepackage{wrapfig}
\usepackage{algorithmic}
\usepackage[framemethod=tikz]{mdframed}
\usepackage{gensymb}
\usepackage[tableposition=below]{caption}
\usepackage[algosection,ruled,vlined,linesnumbered]{algorithm2e}
  \SetKwInput{KwIn}{Inputs}
  \SetKwFunction{Frefine}{Refine}
  \SetKwFunction{Ffilter}{Filter}
  \SetKwFunction{Fviz}{Visualize}
  \SetKwFunction{Fverify}{Verify}
  \SetKwFunction{Fpre}{GetPrecondition}
  \SetKwFunction{Fexplain}{Explain}
  \SetKwFunction{Foracle}{Oracle}
  \SetKwFunction{Fwp}{WeakestPre}
  \SetKwFunction{Fcomp}{Compare}
  \SetKwProg{Fn}{}{:}{}
  \DontPrintSemicolon
\usepackage{stmaryrd}
\usepackage[subtle,paragraphs=normal]{savetrees}
\usepackage{tikz}
\usetikzlibrary{shapes.geometric, arrows,shadows,positioning}
\usetikzlibrary{backgrounds}
\usepackage{enumitem}

\DeclareMathSymbol{\mlq}{\mathord}{operators}{``}
\DeclareMathSymbol{\mrq}{\mathord}{operators}{`'}
\DeclareMathOperator*{\argmin}{arg\,min}

\newcommand{\conspec}{\texttt{Con}_{\texttt{spec}}}
\newcommand{\strength}{\texttt{>}}

\newcommand{\encode}[1]{#1_{enc}}
\newcommand{\head}[1]{#1_{head}}

\newcommand{\Labels}{C}

\newcommand{\repmap}{r_{map}}

\newcommand{\fun}{\rightarrow}

\newcommand{\condir}[1]{\overline{#1}}

\newcommand{\Dtrain}{D_{\text{train}}}

\definecolor{turquoisegreen}{rgb}{0.63, 0.84, 0.71}
\definecolor{tuftsblue}{rgb}{0.28, 0.57, 0.81}
\definecolor{ForestGreen}{RGB}{34,139,34}

\begin{document}

%\title{Leveraging Vision-Language Models for Debugging and Runtime Analysis of Deep Neural Networks\\ {(A Case Study)}}

\title{Debugging and Runtime Analysis of Neural Networks with VLMs {(A Case Study)}}

\author{
\IEEEauthorblockN{Boyue Caroline Hu}
\IEEEauthorblockA{
\textit{University of Toronto, Canada}\\
%Email: boyue@cs.toronto.edu
}
\and
\IEEEauthorblockN{Divya Gopinath}
\IEEEauthorblockA{
\textit{KBR Inc., NASA Ames, USA}\\
%Email: divya.gopinath@nasa.gov
}
\and
\IEEEauthorblockN{Corina S. P\u{a}s\u{a}reanu}
\IEEEauthorblockA{
\textit{KBR Inc., Carnegie Mellon Univ., NASA Ames, USA}\\
%Email: corina.s.pasareanu@nasa.gov
}
\and
\IEEEauthorblockN{Nina Narodytska}
\IEEEauthorblockA{
\textit{VMware Research, USA}\\
%Email: nnarodytska@vmware.com
}
\and
\IEEEauthorblockN{Ravi Mangal}
\IEEEauthorblockA{
\textit{Colorado State University, USA}\\
%Email: ravi.mangal@colostate.edu
}
\and
\IEEEauthorblockN{Susmit Jha}
\IEEEauthorblockA{
\textit{SRI International, USA}\\
%Email: susmit.jha@sri.com
}
}

\maketitle

\begin{abstract}
Debugging of Deep Neural Networks (DNNs), particularly vision models, is very challenging due to the complex and opaque decision-making processes in these networks. In this paper, we explore multi-modal Vision-Language Models (VLMs), such as CLIP, to automatically interpret the opaque representation space of vision models using natural language. This in turn, enables a semantic analysis of model behavior using human-understandable concepts, without requiring costly human annotations. 
%The analysis is summarized in 
Key to our approach is the notion of {\em semantic heatmap}, that succinctly captures the statistical properties of DNNs in terms of the concepts discovered with the VLM and that are computed off-line using a held-out data set.
 We show the utility of semantic heatmaps for fault localization -- an essential step in debugging -- in vision models. Our proposed technique helps localize the fault in the network (encoder vs head) and also highlights the responsible high-level concepts, by leveraging novel {\em differential heatmaps}, which summarize the semantic differences between the correct and incorrect behaviour of the analyzed DNN.  We further propose a lightweight runtime analysis to detect and filter-out defects at runtime, thus improving the reliability of the analyzed DNNs. The runtime analysis works by measuring and comparing the similarity between the heatmap computed for a new (unseen) input and the heatmaps computed a-priori for correct vs incorrect DNN behavior. We consider two types of defects: misclassifications and vulnerabilities to adversarial attacks.
    We demonstrate the debugging and runtime analysis on a case study involving a complex ResNet-based classifier trained on the RIVAL10 dataset. 
    %We consider two types of defects: misclassifications and vulnerabilities to adversarial attacks.
   %Our approach provides intuitive insights into model behavior and thus guides the development of more reliable DNNs for perception tasks. 
\end{abstract}

\begin{IEEEkeywords}
Deep Neural Networks, Debugging, Multi-modal models, VLMs, Fault localization, Runtime Monitoring 
\end{IEEEkeywords}

\section{Introduction}
\label{sec:introduction}

DNNs are increasingly used in many applications, impacting many aspects of our life. Unfortunately. like any other software, DNNs can have bugs. DNNs are notoriously hard to debug due to their complex and
opaque decision-making processes.
There are a number of sophisticated approaches for debugging DNNs
%~\cite{DeepLocalize,DeepDiagnosis,DeepCNN,Kakavandi2023InterpretableFD,NeuraLint,JainLMM23}, 
including spectrum-based~\cite{ma2018mode,eniser2019deepfault}  and mutation-based~\cite{GhanbariTAR23} techniques.
%[CITE ALL EXISTING TECHNIQUES, DEEPFAULT, DEEP LOCALIZE, DEEPMUFL]. 
They focus on locating and correcting bugs in syntactic structural elements of DNNs, such as specific layers or neurons with faults, faulty activation functions, incorrect learning rates so on.
However current techniques do not provide an intuitive, global understanding of the underlying problems, which may prevent developers in quickly identifying the characteristics of the inputs that lead to failures and promptly fixing the discovered faults. Attribution approaches~\cite{simonyan2014deep,sundararajan2017axiomatic,SelvarajuCDVPB20} typically attempt to identify portions of an input that impact model prediction the most. This can help debug individual tests but again does not provide a global view of the (reason for) faulty behaviour.

%which can enable a more semantic reasoning of failures in terms of human-understandable concepts.
%Interpretable Debugging [READ 21, 22] focuses on this aspect, but it does not ... and is more expensive ...WHY?
%In this work, we attempt to present a practical (inexpensive) way to enable interpretable debugging.

To address this problem, we explore
multi-modal Vision-Language Models (VLMs), such as CLIP~\cite{pmlr-v139-radford21a}, to automatically interpret the opaque representation space of  vision models using natural language. VLMs are trained on a large corpus of images accompanied by captions describing the images and are thus aware of high-level, human-understandable concepts describing a variety of images. We leverage VLMs to perform a {\em semantic analysis} of a separate %%\divya{\textit{another} instead of \textit{seperate}?} 
vision model with respect to a set of concepts. Key to our approach is the notion of a {\em semantic heatmap}, that succinctly captures the statistical properties of the analyzed model in terms of the concepts. The VLM and the heatmaps help give {\em semantically meaningful insights} for the understanding and debugging of the the model with respect to a {\em set of inputs}. %\divya{should we give a small example with relevant concepts to illustrate what we mean by semantic analysis?}

We first describe a novel technique for fault localization -- which is a key step in any debugging activity.   Given a set of mis-classified images, the proposed technique uses a map that aligns the model's internal encoding representation with the VLM representation to compare the outputs of the model and the VLM on the same inputs. This helps identifying if the reason for the mis-classification is due to a bug in the encoder of the vision model or in its classification head. We further use semantic heatmaps to validate the location of the bugs and identify the reason for failure in terms of the concepts. To this end, we compute {\em differential heatmaps}, which summarize the semantic differences between the correct and incorrect behaviour of the analyzed DNN and thus pinpoint the problematic concepts.

We also propose a lightweight runtime analysis to detect and filter-out defects at runtime, thus improving the reliability of the analyzed DNNs.
The runtime analysis works by measuring and comparing the similarity between a heatmap computed for a new (unseen) input and the heatmaps computed a-priori for correct vs incorrect DNN behavior. If the heatmap of the input is found to be more similar to the heatmap that summarizes correct behaviour, we deem it as {\em good} and we conjecture that the DNN gives the correct answer for it; otherwise we flag it as problematic. 

We illustrate the proposed techniques on a case study using the  RIVAL10 dataset and a fairly complex image classification ResNet18 model. We study faults that are due to model inaccuracies or adversarial attacks. We provide all the artifacts for our experiments \href{https://anonymous.4open.science/r/concept-heatmaps-3720}{here}.  
%\noindent
%\textit{\textbf{Contributions.}} 

We summarize our contributions as follows:%\corina{this list is too long}
\begin{itemize}
\item \textbf{Novel fault localization}: We present an analysis technique that leverages VLMs to {\it localize\it} errors in a separate vision model, classifying them as {\it encoder\it} or  {\it head \it} errors. The technique leverages novel
{\em differential heatmaps} to isolate predicates in terms of high-level
concepts responsible for errors. The technique can handle different {\em types of errors}, including misclassifications due to model inaccuracies and vulnerabilities to adversarial perturbations. In the latter case, the technique helps identify %finding 
robust and non-robust features in the model.
\item \textbf{Novel runtime defect detection}: We describe a lightweight  technique for detecting misclassifications and adversarial inputs at runtime, by comparing the heatmap of a new input with {\it summary heatmaps\it} of correctly and incorrectly classified inputs -- and thereby improving the robustness and reliability of analyzed models.
\item \textbf{Case study}: We illustrate the proposed techniques on a case study from the image classification domain.  
%We demonstrate through experiments how the fault localization technique can localize different types of errors, while the computed heatmaps highlight the concepts and predicates responsible for errors. 
Our results indicate that most vulnerability errors are due to encoder errors (90\% of adversarial inputs), while for misclassifications this picture is more nuanced (approx. 40\% encoder errors and 60\% header errors). Further, we present results for runtime detection of misclassifications and adversarial perturbations, considering multiple norm-bounded attacks, achieving approx. 80\% detection accuracy for both types of errors.
\end{itemize}

We envision that the fault localization results and semantic heatmaps computed with the help of VLMs can be further leveraged to perform targeted testing, re-training, and repairing of vision DNNs. We leave these activities for future work. We see our approach as complementary to existing approaches
%~\cite{DeepLocalize,ma2018mode,eniser2019deepfault,GhanbariTAR23,DeepDiagnosis,DeepCNN,Kakavandi2023InterpretableFD,NeuraLint,JainLMM23,abs-1811-08425}, all 
contributing towards the larger goal of safety assurance of DNN-enabled systems.

\section{Preliminaries}
\label{sec:background}
%In this section, we review the necessary background on neural networks, vision-language models and the %%$~\conspec$  language. 
% \begin{itemize}
%     \item Neural Networks
%     \item cosine similarity
%     \item VLMs
%     \item $Con_{spec}$
% \end{itemize}

%\corina{from the other paper: needs to be edited}

\vspace{0.05in}
\noindent
\textit{\textbf{Neural Networks.}}
We consider feed-forward neural networks as functions $f:X\rightarrow Y$, where $X$ is a (high-dimensional) space $\mathbb{R}^d$ 
%over real vectors 
and $Y$ is $\mathbb{R}^{|\Labels|}$ where $\Labels$ is a finite set of labels or classes. For classification, the output
defines a score (or probability) across $|\Labels|$; the class with the highest score is output as the prediction.
Previous work~\cite{pmlr-v80-kim18d,Zhou_2018_ECCV,crabbe2022concept,moayeri2023text} indicates that neural classifiers can be seen as the composition of an encoder, denoted here as $\encode{f}:X\rightarrow Z$, and a head, denoted here as a $\head{f}:Z\rightarrow Y$, where $Z$ ( $\mathbb{R}^{d'}$) is the \emph{embedding (or representation) space} of the network. 

The encoder translates the inputs, e.g., pixels forming images, into higher-level representations, and the head computes the appropriate label based on these representations. 
As an example, in a typical convolutional neural network, the convolutional layers act as the encoder and are responsible for extracting high-level 
% susmit: there was a missing word - not sure you want concepts or features here
concepts from the inputs, while the fully-connected layers act as the head, and are responsible for classification based on the extracted concepts. 
%Note that the encoder and the head can each be decomposed into a number of linear and non-linear operations (referred to as \emph{layers}). However, the internal structure of encoders and heads are not relevant for the purpose of this paper, so we treat them as black-boxes unless noted otherwise. 
The  \emph{encoding of an input} $x$ is $\encode{f}(x)$.
%We next define the notion of a \emph{embedding of an input} (also referred to as \emph{representation} or \emph{encoding} of an input). 
%\corina{do we need as definition?}
%\begin{definition}[Embedding of an input]
%\label{def:input_rep}
%Given a neural classifier $f: X\rightarrow Y$ that can be decomposed into an encoder $\encode{f}:X\rightarrow Z$ and a head $\head{f}:Z\rightarrow Y$, $\encode{f}(x)$ is the representation of input $x$. 
%\end{definition}

%\paragraph{\textbf{Cosine Similarity.}}
%Cosine similarity is a measure of similarity between two non-zero vectors. 
%Given two n-dimensional vectors, $a$ and $b$, their cosine similarity is defined as: 
%$\cosine(a,b)=\frac{a \cdot b}{||a|| \, ||b||}= \frac{\sum_i a_i b_i}{\sqrt{\sum_i a_i^2} \cdot \sqrt{\sum_i b_i^2}}$
%Here $a_i$ and $b_i$ denote the $i^{th}$ components of vectors $a$ and $b$, respectively. The resulting similarity ranges from -1 meaning exactly opposite, to 1 meaning exactly the same, with 0 indicating orthogonality.
%\sum_i \frac

\vspace{0.05in}
\noindent
\textit{\textbf{Vision-Language Models.}}  Vision-language models (VLMs)~\cite{pmlr-v139-radford21a}, denoted as $g$, consist of two encoders, one for images ($\encode{g}^{img}:X\rightarrow Z$) and one for text ($\encode{g}^{txt}:T \rightarrow Z$), both mapping to the same representation space $Z$.
VLMs are trained on image-caption pairs such that, for each pair, the representation of the image and corresponding caption are as similar as possible (e.g., measured via cosine similarity $\cos$) in the common %representation 
space $Z$. 
VLMs can be used for a variety of tasks such as image classification, visual question answering, or image captioning.

By default, a VLM can be used to select the caption (from a set of captions) which has the highest similarity with a given image. This strategy can be leveraged for {\em zero-shot classification} of images~\cite{pmlr-v139-radford21a} (this is used to design $\head{g}$ here). For instance, given an image from the ImageNet dataset, the label of each ImageNet class, e.g., \textit{truck}, can be  turned into a caption such as \textit{``An image of a truck''}. Then one can compute the cosine similarity between the embedding of the given image with the text embeddings corresponding to the captions constructed for each class, and pick the class that has the highest similarity score.

\vspace{0.05in}
\noindent
\textit{\textbf{Aligning a vision model with the VLM.}} 
Previous work \cite{moayeri2023text} has shown that one can build an affine map $\repmap$ that links the representation spaces of the vision-based DNN and the VLM, $\repmap:Z_f\fun Z_g$. 
%Following previous work \cite{moayeri2023text}, $\repmap$ can be restricted to the class of affine transformations, i.e., $\repmap(z) := M z + d$, where $\repmap$ is learnt by solving the 
The map has the form $\repmap(z) := M z + d$ and is learned by solving an optimization problem:
\begin{equation}
\label{eqn:opt}
    M, d = \argmin_{M, d} \frac{1}{|\Dtrain|} \sum_{x \in \Dtrain} \|M \encode{f}(x) + d - \encode{g}^{img}(x)\|_2^2.
\end{equation}
More complex maps can be defined and learned as well. We learn and leverage such a map in order to use the representation space of a VLM for interpreting the behavior of a separate vision model.

\vspace{0.05in}
\noindent
\textit{\textbf{Concepts and Predicates.}} We aim to reason about vision models in terms of high-level, human understandable concepts. For instance, for a vision model tasked with distinguishing between several classes such as \textit{cat, dog, bird, car,} and \textit{truck}, the relevant concepts could be \textit{metallic, ears,} and  \textit{wheels}. 
Inferring representations of the concepts learned by a model  can be very challenging, typically requiring  manual annotations of inputs with the presence/absence of relevant concepts~\cite{CrabbeS22,KimWGCWVS18,GhorbaniWZK19,Zhou_2018_ECCV,pmlr-v80-kim18d,DBLP:series/faia/YehKR21}. 
Previous work~\cite{mangal2024conceptbased} has shown how to leverage a separate VLM and the affine map to reason about concepts.  Essentially, this is achieved in a way that is similar to how VLMs perform zero-shot learning, 
%classification
using  embeddings of concepts rather than of classes. While the goal of the work in~\cite{mangal2024conceptbased} was to formally prove properties in terms of the extracted concepts, in this work we aim to use concepts in complementary activities, namely fault localization and runtime defect detection.

Given an image $x$, suppose one is interested in the presence or the absence of concepts $con_1$ and $con_2$ in the representation space of $f$ for the image.
Let $\condir{con}$ denote the mean of the text embeddings (in $g$) for a set of captions  that all refer to a concept $con$ in different ways.  
The mean is computed since there may be many different valid captions constructed for the same concept and each caption may lead to a slightly different embedding. One can then compute the similarity score between the (embedding of the) image (mapped to the VLM) and the (embedding of) each concept and pick the concept that has the highest score, i.e., compute and compare: 
%$\cos(\repmap(\encode{f}^{img}(x)),\condir{con_1)$ 
%with $\cos(\repmap(\encode{f}^{img}(x)),\condir{con_2)$.
$$\cos(\repmap(\encode{f}^{img}(x)),\condir{con_1}) >\cos(\repmap(\encode{f}^{img}(x)),\condir{con_2})$$
Following~\cite{mangal2024conceptbased},  we use predicates of the form $con_1 \strength con_2$ as a shortcut for the above. As an example, for an image representing a \textit{truck}, one may expect that the model is more aware of concepts \textit{metallic} and \textit{wheels} rather than \textit{ears}, written as {\em metallic $\strength$ ears} and {\em wheels $\strength$ ears}, as {\em ears} are typically associated with an animal.

\section{Case Study}
\label{sec:case_study}
We demonstrate the proposed debugging and runtime detection on a case study analyzing a vision model trained on the RIVAL10 dataset~\cite{moayeri2022comprehensive}.
%using a vision model ResNet18~\cite{he2016deep} and the RIVAL10 dataset~\cite{moayeri2022comprehensive}, with CLIP~\cite{pmlr-v139-radford21a} as the VLM\footnote{Implementation of our approach and this case study can be found \textcolor{blue}{\href{https://anonymous.4open.science/r/concept-heatmaps-3720/README.md}{HERE}.}\corina{this should be intro and not a footnote}}. 
%Our experimental setup is similar to previous work~\cite{mangal2024conceptbased}, which showed that CLIP can be used to correctly identify many relevant RIVAL-10 concepts.

%\caroline{TODO: add rival10 classes and concepts in figure 1}
\subsubsection{Dataset and Concepts}
\textbf{RIVAL10} is a subset
%comprising 26,000 images 
of ImageNet~\cite{imagenet}, categorized into 10 classes corresponding to those of CIFAR10, including \textit{truck}, \textit{car}, etc.
%(see Fig.~\ref{fig:framework}). %\textit{plane}, \textit{ship}, \textit{cat}, \textit{dog}, \textit{equine}, \textit{deer}, \textit{frog}, and \textit{bird}. 
%The dataset is divided into a training set with 21,000 images and a test set with 5,000 images.
%Additionally, 
The dataset includes annotations by domain experts, providing instance-wise labels for 18 attributes that reflect human-understandable concepts, including \textit{wheels}, \textit{tall}, etc.
%; there are 18 different concepts for this datset.
%[tbd: to check if 18 is correct and provide a figure with all the possible concepts?]. 
While our approach does not require concept annotations for each image, we use these annotations here to identify the relevant concepts for each class (which would normally be supplied directly by domain experts).
The {\em dictionary} of relevant concepts for each class is provided below:\vspace{+0.2cm}\\
{\em class truck:\{wheels, text, metallic, rectangular, long, tall\};\\
class car: \{wheels, metallic, rectangular, \};\\
class plane: \{wings, metallic, long, tall\};\\
class ship: \{metallic, rectangular, wet, long, tall\};\\
class cat: \{ears, colored-eyes, hairy\};\\
class dog: \{long-snout, floppy-ears, ears, hairy\};\\
class equine: \{long-snout, ears, tail, mane, hairy\};\\
class deer: \{long-snout, horns, ears, hairy\};\\
class frog: \{wet\};\\
class bird: \{wings, beak, patterned\};\\
} \vspace{+0.02cm}
%Figure~\ref{fig:hitmap} illustrates the relevant concepts that are computed for each class; on the y axis we list the classes, on the x axis we list the concepts; green boxes mark the relevant concepts. 

For instance, for class \textit{truck}, concepts  \textit{wheels, tall, long, rectangular, metallic} and \textit{text} are all considered relevant whereas all other concepts (e.g., \textit{tail} and \textit{wings}) are irrelevant. Based on these concepts, we define predicates relating all the possible pairs of concepts (a total of $18\times18=324$ predicates). We also identify the relevant predicates (of the form $relevant \strength irrelvant$) for each class.
%Correspondingly, we can elicit strength predicates between pairs of relevant and irrelevant concepts for each class; 
For instance, for class \textit{truck}, we identify 72 relevant predicates, e.g., $wheels \strength wings$ and $metallic \strength  tail$.

%, from a total of $18\times18=324$ predicates. [corina what is 18 here?]
%It is important to note that, in practice, we expect human domain experts to provide only the concepts relevant to each class. There is no need to manually annotate each image which would be very time-consuming.

%\subsubsection{Assumptions.} [corina: this is not true??] We use the same assumptions as in~\cite{mangal2024conceptbased}. In particular, we require that the VLM has a high satisfaction probability for relevant strength predicates of each RIVAL-10 class while also demonstrating high classification accuracy. We also require that $r_{map}$ is of high-quality, i.e., it does not introduce noise when mapping between the two representation spaces. We validate that these assumptions hold for our case study.

\subsubsection{Vision Model}
\begin{figure}[t]
    \centering
    \includegraphics[width=0.95\linewidth,height=0.25\textwidth]{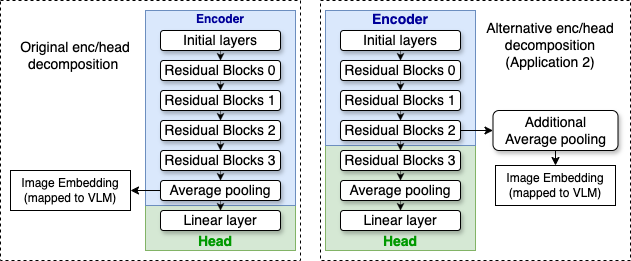}
    
    %\vspace{-0.1in}
    
    \caption{Architecture of the vision model ResNet18. The left side shows the encoder-head decomposition of ResNet18, and the right side shows an alternative decomposition. }
    \label{fig:resnet18-decomp}
    %\vspace{-0.2in}
\end{figure}

We consider a vision classifier, ResNet18~\cite{he2016deep}, trained on the RIVAL10 dataset. %\footnote{\url{https://github.com/mmoayeri/RIVAL10/tree/gh-pages}. The model is pre-trained on the full ImageNet dataset and the final layer of the model, i.e., the head, is further fine-tuned on the RIVAL10 dataset in a supervised fashion using the class labels. \corina{please do not put this in footnotes but in text and give citations}} 
%\textit{Describe the encoder architecture.} 
%As shown on the left side of Fig.~\ref{fig:resnet18-decomp}, 
The architecture of ResNet18 can be decomposed into an encoder and a head for classification based on the encoded images; Fig.~\ref{fig:resnet18-decomp} (left) shows such a decomposition. 
%\corina{[tbd to explain more: why there could be different decompositions]} 
The encoder contains the initial convolutional and pooling layers, followed by four groups of residual blocks consisting of convolutional, batch normalization, and ReLU layers, and then finally an average pooling layer for dimension reduction. The output of the average pooling layer is the image embedding to be mapped to the VLM for analysis. 
The head is a single linear layer with no activation functions with inputs of type $\mathbb{R}^{512}$ and outputs of type $\mathbb{R}^{10}$.

\section{Approach}
\label{sec:overview}

\begin{figure}
\centering
\includegraphics[width=\linewidth,height=0.35\textwidth]{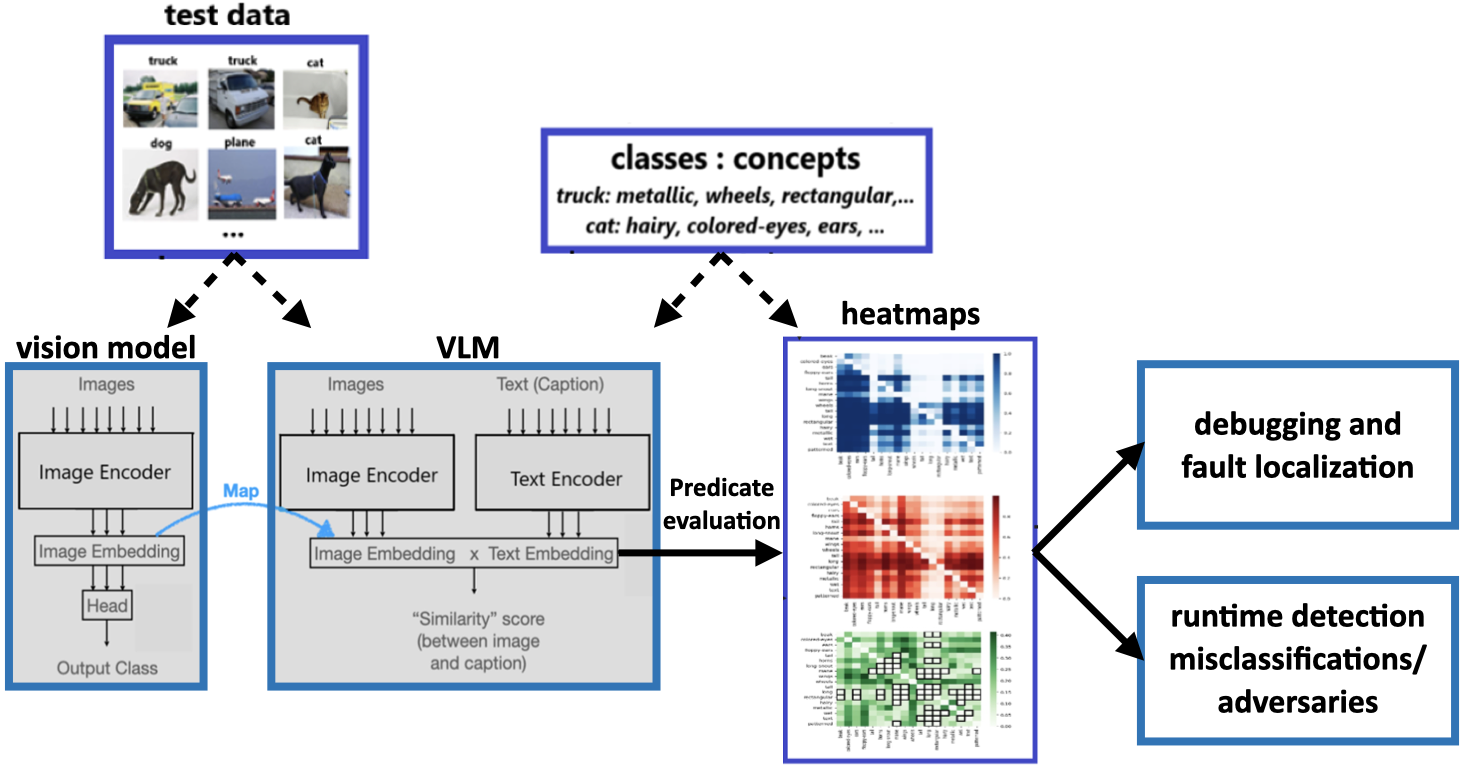}
%\caption{Framework}
\caption{Leveraging VLMs for debugging and run-time analysis of vision models. 
%new figure can be edited \href{https://drive.google.com/file/d/1wG48m9f1Me9qLrMzFk4W1y66U9L7bnpS/view?usp=sharing}{HERE} \nina{Very nice diagram!}
} %\divya{Agree!}
\label{fig:framework}
%\vspace{-0.1in}
\end{figure}

Figure~\ref{fig:framework} provides a high-level overview of our approach to VLM-based debugging and run-time analysis of vision models. Assume we are given a vision model to analyze and a dictionary of high-level concepts that are relevant for each class in the vision task. We also have access to a set of test images (annotated with ground truth for each output class but not concepts). 
%The goal is to analyze, and possibly prevent defects in the vision model.

We use a separate off-the-shelf VLM and build an {\em affine map} to align the embedding spaces of the vision model and the VLM. We leverage the map and the concept definition to define {\em concept predicates} as described in Section~\ref{sec:background}. We evaluate these predicates on the set of test images to produce various {\em semantic heatmaps} that succinctly summarize information about the vision model in terms of high-level, human-understandable concepts. 
The VLM and the semantic heatmaps enable two types of novel analyses: (i) debugging and fault localization and (ii) runtime defect detection. The defects we consider are those that cause misclassifications, and vulnerabilities to adversarial attacks. We provide more details below.

\subsection{Semantic Heatmaps}
%\divya{Do we need few sentences describing the format of our semantic heatmaps? Y-axis and X-axis contain all the concepts. Each cell represents a predicate $con_1 >  con_2$, where $con_1$ is the concept on the Y-axis and $con_2$ is the concept on the X-axis.}
We build semantic heatmaps as 2-dimensional graphs where the X and Y axes contain all the concepts. Each cell represents a predicate $con_1 \strength  con_2$, where $con_1$ is the concept on the Y-axis and $con_2$ is the concept on the X-axis.

\noindent\textit{\textbf{Single-input Heatmaps.}} This simple heatmap is defined per input; every cell in the graph has a value of 0 or 1 corresponding to the valuation of the respective concept predicate. 

\noindent
\textit{\textbf{\textcolor{blue}{Ground-truth Summary Heatmaps.}}}
This heatmap summarizes the satisfaction probabilities for the concept predicates for a set of inputs that {\em all have the same ground truth}. 
%[todo: explain how we compute satisfaction probabilities]. 
Given a set of inputs, the satisfaction probability for a concept predicate ($con_{i}\strength con_{j}$) measures the ratio of the inputs in that set that satisfy the predicate.

%$$Psat_{i,j} =  \frac{\#\{x \mid \;\; x \in I \;\; \land \;\; con_{i} \strength con_{j} (x) = \text{True}\}}{\#I}$$

\begin{figure*}[t]
    \centering
    \begin{subfigure}{0.22\linewidth}
        \centering
        \includegraphics[width=\linewidth]{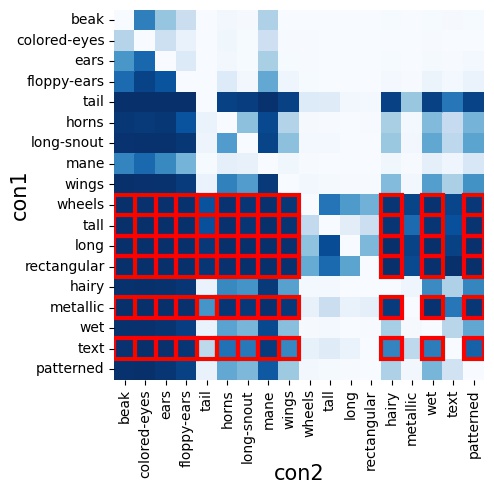}
    %\vspace{-0.25in}
    \caption{With all images with ground truth \textit{truck}.}
    \label{fig:all_truck}
    \end{subfigure}
    \hspace{0.05in}
    \begin{subfigure}{0.22\linewidth}
        \centering
        \includegraphics[width=\linewidth]{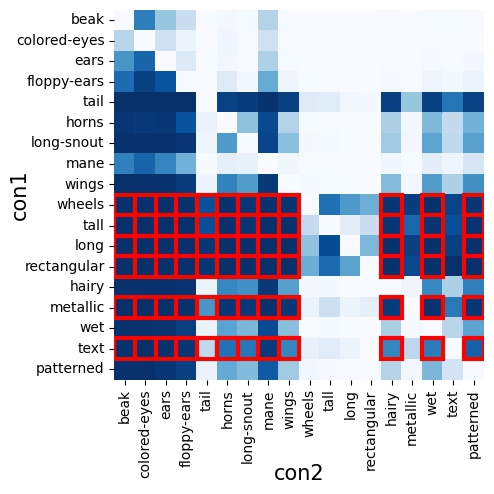}
        \vspace{-0.25in}
        \caption{With images correctly classified as \textit{truck}.}
        \label{sfig:correct_truck}
    \end{subfigure}
    \hspace{0.05in}
    \begin{subfigure}{0.22\linewidth}
        \centering
        \includegraphics[width=\linewidth]{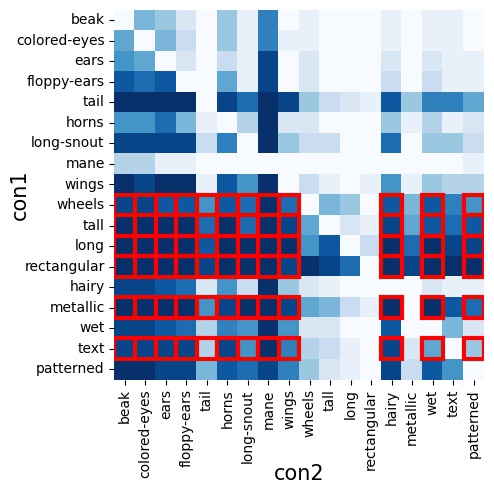}
        \vspace{-0.25in}
        \caption{With misclassified \textit{truck} images due to encoder error.}
        \label{sfig:truck_enc}
    \end{subfigure}
    \hspace{0.05in}
    \begin{subfigure}{0.22\linewidth}
        \centering
        \includegraphics[width=\linewidth]{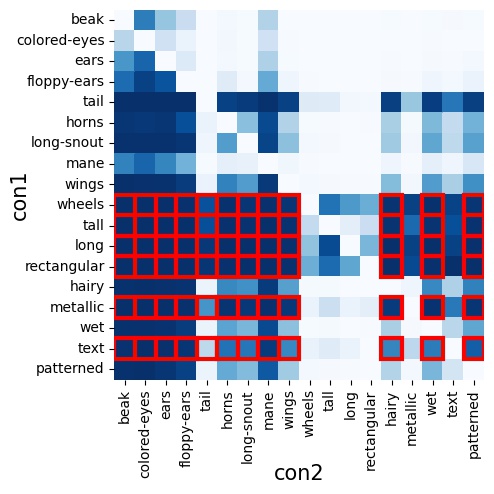}
        \vspace{-0.25in}
        \caption{With misclassified \textit{truck} images due to head error.}
        \label{sfig:truck_head}
    \end{subfigure}
    \begin{subfigure}{0.22\linewidth}
        \centering
        \includegraphics[width=\linewidth]{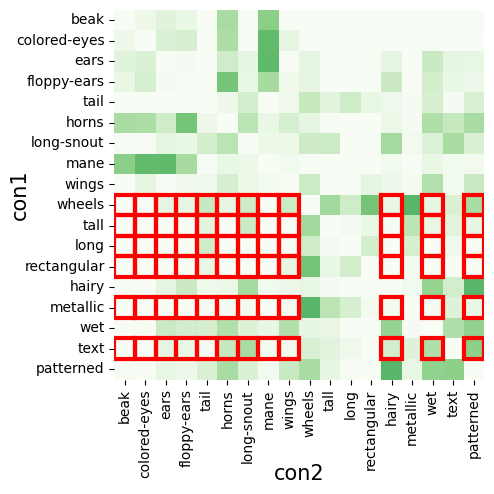}
    %\vspace{-0.25in}
    \caption{Differential heatmap for encoder errors.}
    \label{sfig:diff_enc_truck}
    \end{subfigure}
    \hspace{0.15in}
    \begin{subfigure}{0.22\linewidth}
        \centering
        \includegraphics[width=\linewidth]{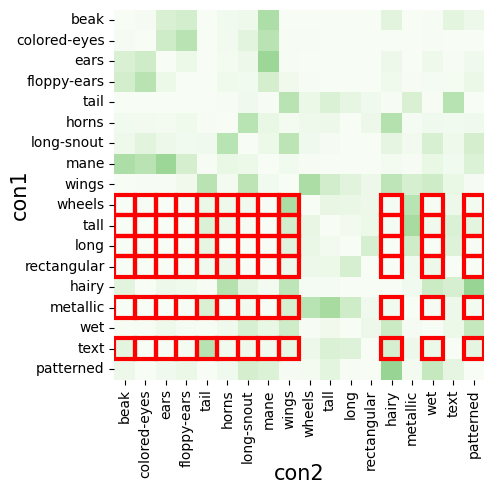}
        \vspace{-0.25in}
        \caption{Differential heatmap for head errors.}
        \label{sfig:diff_head_truck}
    \end{subfigure}
    \vspace{-0.1in}
    \caption{Ground-truth summary and Differential heatmaps for \textit{truck} images. Red outlines indicate relevant predicates for \textit{truck}. }
    \label{fig:resnet-error}
    \vspace{-0.15in}
\end{figure*}

Throughout the paper we depict these heatmaps in blue. For example, Figure~\ref{fig:all_truck} presents the ground-truth heatmap for a set of images with ground-truth label {\em truck}.  
%[tbd explain all heatmaps (a--d) in figure 3]
The dark color corresponding to {\em metallic $\strength$ ears} indicates high satisfaction probability. This is as expected since concept {\em metallic} is relevant for class {\em truck} while concept {\em ears} is irrelevant; relevant and irrelevant concepts for a class are indicated by the user. 

To enable different kinds of analyses,%as described below, 
we compute the same kind of heatmap for different groups of inputs. For example, Figure~\ref{sfig:correct_truck} presents the ground-truth heatmap for the sub-set of images with ground-truth label {\em truck} which are all {\em correctly classified} by the vision model, while Figures~\ref{sfig:truck_enc} and  ~\ref{sfig:truck_head} present the ground-truth heatmaps for sub-sets of images with ground-truth label {\em truck} which are {\em misclassified}, due to different errors.

Throughout the paper we depict predicates that relate {\em relevant and irrelevant concepts} for a particular class with a red outline in the different heatmaps.

\noindent
\textit{\textbf{\textcolor{orange} {Output-label Summary Heatmaps.}}}
This heatmap is similar to the previous one, but it summarizes the satisfaction probabilities for the predicates for a set of inputs that {\em all have the same output label} (but may have different ground-truth).  Throughout the paper we depict these heatmaps in orange (see Figure 7a). %Fig~\ref{sfig:detect_pgd} presents the output-label heatmap for a set of images which are classified as {\em truck} (they may or may not have ground truth {\em truck}). 
%\corina{fix figure and comment on the subtle differences between this heatmap and the blue heatmap}

%Given a set of inputs classified by a model to the same label, we would like to determine if the relevant strength predicates for the output label are satisfied in the representation space, providing a means to justify the classification provided by the model. The output-label based summary heatmap serves to address this by presenting a summarized view of the valuation of the $\conspec$ predicates corresponding to a set of inputs having the same model output (label).  %For example, \textbf{Fig TBD} presents the Output-label based Summary Heatmap for all images classified as truck by the vision model. The red outline represents {\it relevant strength predicates \it} for {\it truck \it}.

\noindent
\textit{\textbf{\textcolor{ForestGreen}{Differential Heatmaps.}}}
This type of heatmap enables a differential analysis between different categories of inputs. 
Given two heatmaps, $H_1$ and $H_2$, of the same dimensionality, a differential heatmap is constructed by calculating the magnitude of the difference between the values in each cell. 
$\forall i,j \;\; H_{\mathit{diff}}[i,j] = | H_1[i,j] - H_2[i,j] | $. The higher the value in a cell, the more the two heatmaps differ in the satisfaction of the corresponding predicate. The darkness of the color in each cell is proportional to the difference. Note that the heatmaps being compared could be of any type (single-input, ground-truth or output-label  heatmaps). Throughout the paper we depict these
heatmaps in green (see Figure~\ref{sfig:diff_enc_truck} and Figure~\ref{sfig:diff_head_truck} for examples). 

\noindent
\textit{\textbf{{Binarized Heatmaps}.}} For defect detection, we also find it useful to {\em binarize} the summary heatmaps, with a threshold $t$, 
%to highlight the strength predicates with high satisfaction probabilities, 
i.e., 1 if the satisfaction probability  is $\geq t$, and $0$ otherwise. 

%\textbf{Fig TBD} shows an example of a differential heatmp between ...

%\susmit{We need to take a few examples and use it across Section 3 to explain these otherwise it becomes very dense reading. Readers will get it rightaway from examples.}
%\corina{if we have space repeat figures 3a, 7a and 6b here and make them bigger}

%\noindent{Note.} We refer to predicates that relate relevant and irrelevant concepts for a particular class 
%and have high satisfaction probability, 
%as {\em relevant strength predicates} for that class. We depict them with a red outline in the heatmaps.

%,ilyas2019adversarial,zhou2021towards}

%\divya{ Discuss how detection of robust features and adversarial inputs is enabled.}

%\divya{I am here.}

%\begin{figure}[t]
%    \centering
%    \begin{tabular}{c c}
%        \begin{subfigure}{0.48\linewidth}
%    \centering
%        \includegraphics[width=\linewidth]{heatmaps/clip/concept_clip_truck.jpg}
%    \caption{w.r.t. concepts}
%    \label{fig:clip_con}
%    \end{subfigure} & \begin{subfigure}{0.43\linewidth}
%    \centering
%        \raisebox{0.12\width}{\includegraphics[width=\linewidth]{heatmaps/clip/class_clip_truck.jpg}}
%    \caption{w.r.t. classes}
%    \label{fig:clip_class}
%    \end{subfigure} \\
%    \end{tabular}
%    \caption{Heatmaps of CLIP embedding of images classified as trucks}
%    \label{fig:clip_heat}
%\end{figure}

\subsection{Debugging and Fault Localization}

\vspace{0.05in}
\noindent
\textit{\textbf{Debugging Misclassifications.}}
%Neural networks are hard to debug due to their opaque nature.
%It is a challenge to debug why a model fails on a given set of inputs, given the uninterpretable logic. The first step in debugging is {\it localizing where the fault is located \it}. Given that a vision model can typically be decomposed into an encoder and a head with different functionalities, 
We propose a novel technique that uses VLMs to perform fault localization, which is a key first step in  any debugging task.  Specifically the technique can determine if a wrong classification is due to a bug in the encoder or the head of a vision model, while considering different decompositions (head vs. encoder) for the model.
%\divya{What does "while considering different decompositions" mean here? Our localization approach localizes to encoder or head , given a specific decomposition. The technique does not consider different decompositions.} [??] 
%Specifically, we investigate if there are errors in the representation created by the encoder for a misclassified input, if so, we categorize it as a bug in the encoder, if not, the bug is isolated to be in the head. \divya{Should we describe the localization approach here?} 
Further, by determining which concept predicates are violated by misclassified inputs, we can also {\it localize the reasons for failure\it}, i.e., determine which concepts are poorly understood by the model. This information can in turn be used to improve the model, e.g., by selecting images with the desired concepts and using them for re-training.
%The details are presented in section~\ref{subsec:debugging}.

\subsubsection{Head vs. encoder errors}
Consider an input $x$ with ground truth class $c$  such that $f(x)\neq c$, i.e., it is misclassified by the vision model. 
%The decomposition of the vision model into an image encoder $f_{enc}$ and a classification head $f_{head}$ (see Sec.~\ref{sec:overview}) allows us to localize errors to a specific part of the model. 
The misclassification can be caused by errors in the encoder or in the head.
%, i.e., the encoder does not correctly encode relevant concepts; or errors in the head, i.e., concepts are encoded correctly but the wrong output label is produced; or both. 
%The location of the error can be detected through checking concept predicates. Intuitively, if the predicates relevant for the ground truth label of $x$ are not satisfied by its image embedding, then the encoder has incorrectly encoded the image; otherwise, the error is in the head. %Note that with encoder errors, errors in the head may be present at the same time. However, since updating the encoder would also affect the head, the encoder errors should be addressed first, thus we consider such a case as encoder errors. 
Note that, for an image $x$, encoder and head errors may be present at the same time. However, since updating the encoder would also require updating the head, the encoder errors should be addressed first, thus we consider such cases as exclusively encoder errors. %\ravi{reworded slightly; check}

Assume that $x$ is correctly classified by the VLM via zero-shot classification, i.e., $g_{head}(g_{enc}^{img} (x)) = c$. Further assume $r_{map}$ is of high quality.
%but the model misclassifies, i.e., $f(x) \neq c_0$.
Let $e_1 := r_{map}(f_{enc}(x))$ refer to the embedding of $x$ produced by the vision model and mapped to the VLM space, and let $e_2:= g^{img}_{enc}(x)$ refer to the embedding of $x$ directly produced by the VLM. 

\begin{itemize}

\item{\textit{{Encoder Error.}}}
If $e_1$ and $e_2$ lead to different zero-shot classification outputs, i.e., $g_{head}(e_1) \neq g_{head}(e_2)$, then it is likely that the embedding $e_1$ is incorrect, and hence the error is in $f_{enc}$. We call it \textit{encoder error}. 
%[this is because... to explain]
%\divya{should we mention we assume $g_{head}(e2)$ is correct?}\caroline{added}. 
%\end{definition}

%\begin{definition}[Head Error]
\item{\textit{{Head Error.}}}
%Otherwise, $g_{head}(e_1) = g_{head}(e_2) = c_0 \neq f(x)$, suggesting that the embedding computed by $f_{enc}$ and mapped to the VLM's representation space via $r_{map}$ allows the VLM head to output the correct label, thus the error is in $f_{head}$; we call it \textit{head error}.
If $e_1$ and $e_2$ lead to the same zero-shot classification outputs, i.e., $g_{head}(e_1) = g_{head}(e_2)$, it is likely that the embedding computed by $f_{enc}$ and mapped to the VLM's representation space via $r_{map}$ is correct since it allows the VLM head to output the correct label. Hence the error is in $f_{head}$; we call it \textit{head error}.
%\end{definition}
\end{itemize}

%\nina{If we have space, I would frame these as definitions (or highlighted somehow) as these are basic notions: encoder and head errors.}. %\caroline{added some sentences about Fig.~\ref{fig:enc_head_errs}} %\divya{maybe we could move these details to 4.3.2 , where we refer to Fig 4? }\caroline{I feel like these examples help with understanding the definition of encoder/head error}

%\divya{Should we mention that this 
%\corina{maybe move this to future work}The localization results could potentially be used to selectively re-train or repair either the encoder or head of the model and could be a starting point for more precise layer-wise analysis.

\subsubsection{Analyzing Errors with Heatmaps}
We compute differential heatmaps to help localize the predicates that distinguish between correct vs incorrect inputs, for different types of error (head vs encoder).
Specifically, given the sets of misclassified inputs for a specific ground-truth label categorized as encoder and head errors respectively, we construct ground-truth summary heatmaps for each set. We then compare these heatmaps with the ground-truth summary heatmap for the images correctly classified to the specific label. We do this by constructing a differential heatmap which calculates the absolute difference between the summary heatmap for encoder errors and correctly classified inputs, and a similar differential heatmap for head errors; see Figure~\ref{sfig:diff_enc_truck} and Figure~\ref{sfig:diff_head_truck}. A large difference indicates predicates that distinguish between correct versus incorrect classifications. Potentially, these predicates can be leveraged to retrieve more images for retraining the model.
%[tbd: for example: pick an example from the figure]

\vspace{0.05in}
\noindent
\textit{\textbf{Debugging Vulnerabilities and Identifying Robust and Non-Robust Features.}}
%\divya{Should the heading be more specific like Debugging Adversarial Vulnerability, to differentiate it from the previous section? Mis-classification (covered in the previous subsection) is also a vulnerability of the model.} [\corina{how are misclassifications vulnerabilities??}]
%: Detection of Robust/Non-robust Features.}} 
%[this item is too long compared to the others]
We also consider another type of defect, namely vulnerabilities to adversarial perturbations. Vision DNNs are known to be vulnerable to \textit{adversarial perturbations}; small changes (imperceptible to human eye) applied to validly classified images can fool the model into misclassification~\cite{SzegedyZSBEGF13,Goodfellow2014ExplainingAH,KurakinGB16}. %\corina{[todo: need citations here]}. 
We can apply the same debugging approach for this type of defects.
%Although, most earlier work has typically attributed adversarial vulnerability to the high dimensional nature of the input space or statistical fluctuations in the training data, more 
Recent work ~\cite{ilyas2019adversarial} points out that adversarial examples can be attributed to the presence of brittle {\it non-robust \it} features. 
%These include irrelevant features that the model considers important for a particular class.
%There may also be features relevant to a particular class that the model does not encode in a robust manner; small changes in inputs leads to the feature not being identified. 
%Further, other  related work~\cite{zhou2021towards} suggests that adversarial inputs retain features that humans typically consider for a classification, while additionally adding noise that forces the given model into mis-classification. 
%Such features termed as {\it attack invariant features \it} are {\it robust \it} to adversarial perturbations. 
%
We propose the use of semantic heatmaps to identify robust and non-robust features of a model, where we consider as features the human-understandable concepts as studied in this paper. This is another debugging activity, where the cause of misclassification is potentially the feature (or concept) found to be non-robust. 
%[corina: how is this different from above?]
%[Corina: maybe explain how this is also a form of debugging/fault localization]
%Note that each concept predicate captures a relation between two concepts as encoded in the representation space, which can be considered as a {\it feature \it} of the vision model. %\ravi{to check if we use this definition of feature later}
%\ravi{I replaced all such references to features with strength predicates, so we can safely remove this last sentence to avoid overloading the word feature.}

\subsection{Runtime Defect Detection}

%\vspace{0.05in}
\noindent
%\textit{\textbf{Runtime Detection of Misclassifications.}}
%[tbd]

%\vspace{0.05in}
%\noindent
%\textit{\textbf{Runtime Detection of Adversarial Inputs.}}
We also propose the use of semantic heatmaps for runtime detection of adversarial inputs or of inputs that lead to misclassifications. This can potentially improve model robustness and reliability without incurring the cost of retraining. %[Corina: maybe give the high level idea] 
%The runtime detection works  by comparing the single-input heatmap for the input under consideration with summary heatmaps
%capturing the feature space of 
%capturing the characteristics of adversarial and valid inputs.% (section~\ref{subsec:adv_detect}).
%If the single-input heatmap is judged to be closer to the summary heatmap of valid inputs, it is judged to be benign; but if it is closer to the summary heatmap of adversarial inputs, it is flagged as potentially adversarial.

%[ moved from experiments section]
\noindent\textit{\textbf{Run-time detection of adversarial inputs.}}
For each output class, we construct a ground-truth summary heatmap with all clean (i.e. unperturbed), correctly classified images (as in Figure~\ref{sfig:correct_truck} for \textit{truck}) 
%[is this clean all ie Fig3a or clean correctly classified Fig3b?? im putting back ravi's comment to discuss]
%\ravi{Do we conside ground-truth heatmaps or output heatmaps for clean images? Seems like we should consider output heatmaps} \divya{We consider clean images that are correctly classified, so the ground-truth and output summary heatmaps are the same. But yes, we can mention output heatmap to avoid any confusion.}
and an output summary heatmap with all adversarial inputs misclassified to the class (generated during offline analysis).
%(i.e., a perturbed heatmap as in Fig.~\ref{sfig:detect_pgd} for \textit{truck}). 
Then, for each input encountered during runtime with output class $c$, this input is considered adversarial if its single-input heatmap is more similar to the perturbed heatmap than the clean heatmap of class $c$, and clean otherwise. To determine the similarity between a single-input heatmap and the above summary heatmaps,
%\susmit{is summary over all of training data - different inputs have different heatmaps}\caroline{addressed}, 
we first binarize the summary heatmap with a threshold $t$ %\susmit{some insight into selection of t would be useful}\caroline{added after this sentence, please check} 
to highlight strength predicates with high satisfaction probabilities, i.e., $1$ if satisfaction probability $\geq t$, and $0$ otherwise. 
%The value of $t$ should be sufficiently high to capture the primary differences between clean and perturbed heatmaps while avoiding the inclusion of small values that are predicates satisfied by only a small set of images. In our experiment, we set $t=0.6$, which resulted in the highest detection accuracy. The binarized clean heatmap (binarized Fig.~\ref{sfig:correct_truck}) is shown in Fig.~\ref{sfig:detect_c}; and the binarized perturbed heatmap (binarized Fig.~\ref{sfig:detect_pgd}) is shown in Fig.~\ref{sfig:detect_pgd_b}, with $t=0.6$. 
We compute Intersection over Union (IoU) as the similarity metric, between each single-input heatmap and the binarized summary heatmaps; we use $IoU_p$ for similarity with the the binarized perturbed heatmap, and $IoU_n$ for similarity with the binarized clean heatmap. An input with $IoU_p > IoU_n$ suggests it is adversarial, and clean otherwise.

\noindent \textit{\textbf{Run-time detection of misclassifications.}}
 Even if the input is not adversarially or naturally perturbed, it may lead to misclassifications (due to inaccuracies in the model). Detecting (and possibly preventing) such inputs would improve the reliability of the model, particularly relevant in safety-critical settings. The problem is challenging 
 since the ground-truth is not known at runtime. To detect misclassifications in the absence of ground truth, we can take a similar approach as above, by comparing the single-input heatmap with the summary heatmaps computed for correct vs incorrect behavior.

\section{Experiments}
\subsection{Experimental Setup}
%[tbd]
We demonstrate our techniques on the RIVAL10 case study. The RIVAL10 dataset is divided into a training set with 21,000 images and a test set with 5,000 images.  The ResNet18 model was pre-trained on the full ImageNet dataset and the final layer of the model, i.e., the head, was further fine-tuned on the RIVAL10 training set in a supervised fashion.
The model has an accuracy of $96.73\%$ on the test set. 
\subsubsection{Vision-Language Model}
%\divya{Give details about CLIP.}
As mentioned, we use a pre-trained CLIP model
%\footnote{ \url{https://github.com/openai/CLIP} \corina{change this to a propoer citation}} 
for our experiments, which has a Vision Transformer (ViT-B/16)~\cite{dosovitskiy2021an} as the image encoder and a Transformer\cite{vaswani2017attention} as the text encoder. 
The representation space of the CLIP model is of type $\mathbb{R}^{512}$. The head  $g_{head}$ is a zero-shot classifier, implemented as detailed in Section~\ref{sec:background}. This classifier achieved an accuracy of 98.79\% on the RIVAL10 test set. 
Our experimental setup is similar to previous work~\cite{mangal2024conceptbased}, which showed that CLIP can be used to correctly identify many relevant RIVAL10 concepts.

\subsubsection{Map between ResNet18 and CLIP}
%\divya{Reword and shorten}
The map is obtained by learning an affine map $r_{map}$ from the representation space of the ResNet18 vision model to the representation space of the CLIP model~\cite{moayeri2023text}. We learn a map of type $\mathbb{R}^{512}\fun\mathbb{R}^{512}$, with a low Mean Squared Error (MSE) of 0.963 and a high Coefficient of Determination (R$^2$) of 0.786 on the test data, suggesting good quality.
The problem is solved via gradient descent on RIVAL10 training data for 50 epochs using a learning rate of 0.01, momentum of 0.9, and weight decay of 0.0005. 

\subsection{Debugging Misclassifications}
%and Fault Localization}
\label{subsec:debugging}
%For inputs that are misclassified by the vision model, %i.e., the model output does not match the ground truth, 
ResNet18 misclassifies 163 
%\corina{should be 163??} 
images from the RIVAL10 test set. We investigate the cause of failure by locating if the bug lies in the encoder or in the classification head, and  isolating the predicates that get violated by the misclassified inputs. 

%We investigate the following types of failures; mis-classified in-distribution inputs and inputs with specific types of perturbations. (this is probably not this application)
%{\it How does CLIP perform when fed directly with perturbed inputs? \it} 

%\paragraph{Setup} 
%\begin{itemize}
%    \item Dataset: Rival10
%    \item vision models: Resnet18, Resnet50, vit\_tiny, VLM: clip
%    \item t2c mappings
%    \item perturbations: adversarial, poison, natural (brightness, noise, etc.)
%\end{itemize}

\subsubsection{Localization to Encoder vs Head Errors}
%\subsubsection{Encoder and Head Errors} %\caroline{summary of this paragraph: def of encoder vs head error.}
%\divya{The description of the localization algorithm, given below, could perhaps be moved to the overview section. We can present the results here.}

%\divya{Results start from here.}
%ResNet18 mis-classifies 173 images from the RIVAL-10 test dataset. 
We apply our localization approach to categorize the misclassifications as encoder errors and head errors. Essentially the approach treats the CLIP model as an oracle to validate the %encoding of 
ResNet18 model. Therefore, it requires that the CLIP model classifies accurately when fed directly with the images. Out of the 163 images, 145 images are correctly classified by CLIP. We use these images to localize the errors in the ResNet18 model. Out of these images, we found that 61 images are misclassified due to encoder errors and 84 images due to head errors, using the procedure described in Section~\ref{sec:overview}.

\begin{figure*}
\centering
\includegraphics[scale=0.35]{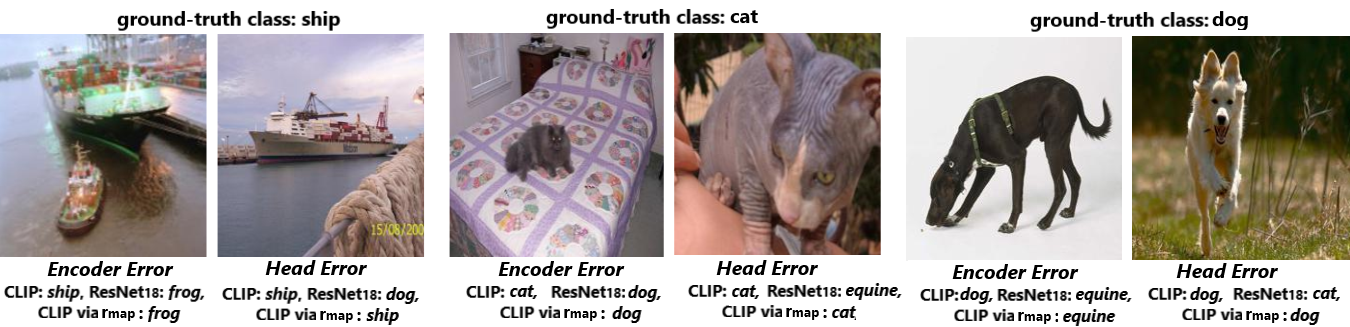}
%\caption{Examples of misclassified inputs for ResNet18 model, localized to bug in the encoder (Encoder Error) and bug in the head (Head Error).}

\caption{Examples of misclassified inputs of ResNet18, error localized to the encoder (encoder error) and the head (head error).}
\label{fig:enc_head_errs}

\end{figure*}

As an example, consider the \textit{ship} images on the left of Figure~\ref{fig:enc_head_errs} that are misclassified by the ResNet18 model.
%ResNet18. 
The one on the left is misclassified due to an encoder error because $g_{head}(e_2)$ (i.e., CLIP) is \textit{ship} but $g_{head}(e_1)$ (i.e., CLIP via $r_{map}$) is \textit{frog}. 
The one on the right is misclassified due to a head error because $g_{head}(e_2)$ and $g_{head}(e_1)$ (i.e., CLIP via $r_{map}$) are both \textit{ship} but the network output is \textit{dog}.

\subsubsection{Validating Fault Localization with Mutations} 
%\subsubsection{Error Localization} %\caroline{summary of this paragraph: we use a mutation-based approach to demonstrate encoder and head error. mutation in encoder [resp. head] led to a significant increase in No. encoder [resp. head] errors. Additionally, with an alternative decomposition, we can localize errors even more, e.g., with residual block 3. }

%\ravi{Section4.3.1 is also about error localization, so I renamed this subsubsection}

%Application of the above technique on the test inputs with ground-truth truck that are misclassified by ResNet18, we categorize 61 as encoder errors and 84 as head errors respectively.

%We would like to use ground-truth based summary heatmaps to validate this categorization. We expect that for inputs localized as encoder errors, the corresponding embedding vectors would not satisfy the strength predicates relevant to the ground-truth label, while the embedding vectors for the head errors would satisfy the relevant strength predicates.

%Show the corresponding heatmaps for the 61 errors and 84 head errors. PUT pictures for ENC and HEAD errors.
To validate our the fault localization technique, we employ a mutation-based approach by manually introducing errors in the encoder or the head of ResNet18: this is achieved by randomizing the weights of a small set of neurons. If our technique is valid, then by introducing errors in the encoder, the localization should result in statistically more encoder errors than head errors, and vice versa.
%\nina{It feels to me that we need to explain what exactly we expect to evaluate using mutations.} 
From ResNet18, we created two different mutated models, one with randomized weights for two neurons in each convolutional layer in the encoder, and one with randomized weights introduced only in the final classification layer, i.e., in the head. From row 2 of Tbl.~\ref{tab:enc_head_errors}, we can see that introducing mutations in the encoder resulted in significantly more encoder errors than head errors, as expected. Similarly, from row 3, mutations in the head resulted in significantly more head errors than encoder errors, again as expected. 
%From row 2 of Tbl.~\ref{tab:enc_head_errors}, we can see that compared to row 1 for the original ResNet18, the number of encoder errors, identified by our localization technique, significantly increased \divya{compared to what?}\caroline{added} after introducing mutations in the encoder. Similarly, shown in row 3 of Tbl.~\ref{tab:enc_head_errors}, head errors identified by our localization technique significantly increased after mutations in the head. 
%\divya{If I understand correctly, we want to say that in row 2 , our technique identifies more encoder errors than head errors which is in alignment with the ground-truth since the mutation is only in the encoder. It is the reverse in row 3, which is also in alignment with the ground-truth. Is this correct?} \caroline{yes, I reworded these sentences a bit}
This suggests that our localization approach, although not perfect, can provide useful insights into localizing the errors.
%either in the encoder or the head.%}the number of encoder and head errors can indicate the location of errors in the vision model. 
% \caroline{discuss the reason for increasing head error when encoder is mutated and the other way. head is dependent on an encoder? more fine-grained localization is required?}

We also attempted to determine if our %localization
approach can perform a more precise layer-wise localization. %A more precise localization can be conducted with two different encoder head decompositions of the same vision model. 
For example, to localize errors specifically in Residual Block 3 as shown in Figure~\ref{fig:resnet18-decomp}, we can consider an alternative decomposition for ResNet18 by moving Residual Block 3 from the encoder to the head. %Now, the encoder contains the initial layers, Residual Blocks 0, 1, 2, and the head contains layers from Residual Block 3 on. 
In this case, mutations introduced specifically in Residual Block 3 would be encoder errors with the original decomposition, but head errors with the alternative decomposition. 
Since all Residual Blocks contain convolutional layers, to obtain an accurate $r_{map}$, we added average pooling to reduce the dimensionality of the output of Residual Block 2 before the mapping to CLIP embedding space\footnote{The resulting mapping has MSE of 1.128 and $R^2$ of 0.749, comparable with the mapping obtained for the original decomposition}. %\caroline{should we discuss limitations of mapping to CLIP somewhere earlier (like background?)} 

The bottom row of Tbl.~\ref{tab:enc_head_errors} presents the number of encoder and head errors with the original and alternate decomposition. Surprisingly, we observe that regardless of the location of the mutation, i.e., mutation in the encoder or the head, the number of head errors exceeds encoder errors. At the same time, the original decomposition yields more encoder errors compared to the alternate decomposition, indicating a shift in the location of the mutation. This underscores the potential of employing different encoder-head decompositions for precise layer-wise error localization; we leave this exploration for future work. %Nonetheless, we leave a more comprehensive evaluation addressing the increase in head errors as future work.

\begin{table}[t]
    \centering
    \caption{Number of encoder and head errors found through error localization for all the images in the test set. }
    \label{tab:enc_head_errors}
    \scalebox{0.82}{
    \begin{tabular}{|c|>{\columncolor{tuftsblue!15}}c|>{\columncolor{turquoisegreen!20}}c|}
        \hline
        %\multirow{ 2}{*}{Mutation location}
        %& \multicolumn{2}{|c|}{Original decomposition} & \multicolumn{2}{|c}{Alternative decomposition} \\
        %\cline{2-5}
        Mutation location& \# encoder error & \# head error \\%& \# enc error & \# head error\\
         \hline
         No mutation (original ResNet18) & 61 & 84 \\%& 80 & 65 \\
         \hline
         Mutation in Encoder & 4271 & 405 \\%& 3361 & 1305\\
         \hline
         Mutation in  Head & 101 & 4571 \\%& 525 & 4147\\
         \hline
         Mutation in Residual Block 3 & \shortstack{1183 (orig decomp)\\ 438 (alt decomp)} & \shortstack{3064 (orig decomp)\\ 3809 (alt decomp)}  \\%& 438 & 3809 \\
         \hline
    \end{tabular}}
    
\end{table}

\subsubsection{Analysis of Errors with Heatmaps} 
%% from old paper
With semantic heatmaps, we can visualize and validate localization results and understand the reasons for failure. We first perform an analysis using the heatmaps for inputs that are correctly classified to validate that the behavior of the vision model can indeed be analyzed semantically in terms of domain-specific concepts.

We use images and concepts related to class \textit{truck} to illustrate. The RIVAL10 test dataset has 502 images with ground-truth \textit{truck}. The ResNet18 model classifies 474 correctly as \textit{truck}. Figure~\ref{fig:all_truck} shows the ground-truth summary heatmap of all images with ground-truth \textit{truck} and Figure~\ref{sfig:correct_truck} 
depicts the heatmap computed with all \textit{truck} images that are correctly classified by ResNet18. We can observe that both heatmaps are visually similar to each other in terms of the satisfaction probabilities of the concept predicates. 
We can observe from  Figure~\ref{sfig:correct_truck} most relevant predicates have high satisfaction probabilities, i.e., the ones involving concepts relevant to \textit{trucks} (highlighted with red outlines), except for the predicate {\em metallic $\strength$ tail}, and predicates involving concept \textit{text}, e.g., {\em text$\strength$ tail} and {\em text $\strength$ horns}. 
This suggests that when ResNet18 correctly classifies images as \textit{trucks}, it is indeed able to recognize the strong presence of concepts that are relevant for that class, such as \textit{long, rectangular, tall, metallic} and \textit{wheels} in these images. However the model struggles to effectively identify \textit{text} in images, and distinguish \textit{metallic} and \textit{tails}. Additionally, we observe that some non-relevant predicates also have high satisfaction probabilities, e.g., {\em tail $\strength$ beak}. This could be the result of other concepts in the images, and the dependency between relevant and non-relevant concepts. %, or how the CLIP model is trained. Further analysis is required to determine the cause, we leave this as future work.
This analysis highlights how the respective heatmaps could be used to semantically explain and validate the behavior of the model.

We proceed to use the heatmaps to specifically investigate the misclassified inputs due to encoder and head errors to understand the reasons for failure semantically in terms of high-level concepts. There are 28 inputs with ground-truth \textit{truck} that are mis-classified to other labels, out of which 22 are correctly classified by CLIP. Our localization technique categorizes 8 of these mis-classifications as encoder errors and 14 as head errors. Figures~\ref{sfig:correct_truck}, \ref{sfig:truck_enc}, and \ref{sfig:truck_head} depict the ground-truth summary heatmaps for images correctly classified as truck, misclassified due to encoder errors, and to head errors, respectively. Figures~\ref{sfig:diff_enc_truck} and ~\ref{sfig:diff_head_truck} depict the differential heatmaps (between the heatmaps computed for correctly classified vs misclassified inputs) for the encoder and head errors, respectively. 
We found that the average difference value across all 324 strength predicates in Figure~\ref{sfig:diff_enc_truck} was 0.115 (120 predicates with values $> 0.115$ ), while for Figure~\ref{sfig:diff_head_truck} the average difference value was smaller 0.077 (102 predicates with values $> 0.077$). Intuitively, this is because heatmaps summarize the satisfaction probability of concept predicates by the encoder's output. Thus, for Figure~\ref{sfig:diff_enc_truck}, the error is in the encoder, so more predicates are violated, while for Figure~\ref{sfig:diff_head_truck}, the error is in the head, so fewer predicates are violated.

%This indicates that the heatmap for encoder errors was more different from the correctly classified heatmap than the heatmap for head errors, thereby validating the categorization performed by our localization technique. %Further, more number of predicates in the encoder errors heatmap (120) have their difference values greater than the mean, as compared to the head errors heatmap (102). 

Focusing on the 72 predicates relevant to class \textit{truck} (highlighted by the red boxes in the figures), we find that the differential heatmap for head errors 
%around 30\% of these predicates 
shows zero or very small differences; only  
%a difference of zero and only 
two predicates show a 
%\corina{to rewrite and remove 'significant'} 
difference greater than 0.2, {\em wheels} $\strength$ {\em wings}  and {\em text} $\strength$ {\em tail}. 
On the other hand, in the differential heatmap for encoder errors, many more predicates, namely {\em wheels $\strength$ tail, wheels $\strength$ long-snout, wheels $\strength$ wings, wheels $\strength$ patterned, text $\strength$ horns, text $\strength$ long-snout, text $\strength$ wet, text $\strength$ patterned, tall $\strength$ long-snout, long $\strength$ tail}, exhibit relatively large differences, ranging from 0.25 to 0.45. 
The results seem to indicate that the model is specifically faulty when processing images containing {\em wheels} or {\em text}, as these concepts frequently appear in the above predicates. Note also that some other predicates such as those comparing the concept {\em mane} with {\em beak}, {\em colored-eyes} and {\em ears} (which seemingly are irrelevant to class \textit{truck}) also exhibit large values in the differential heatmaps, indicating that the model's output is potentially impacted by unintended features.  

Isolation of the concepts and predicates potentially responsible for the failures can help with generating new inputs that satisfy the respective predicates for further debugging, repair and re-training of the model.

\begin{figure*}[t]
    \centering
    \begin{subfigure}[t]{0.22\linewidth}
        \centering
        \includegraphics[width=\linewidth]{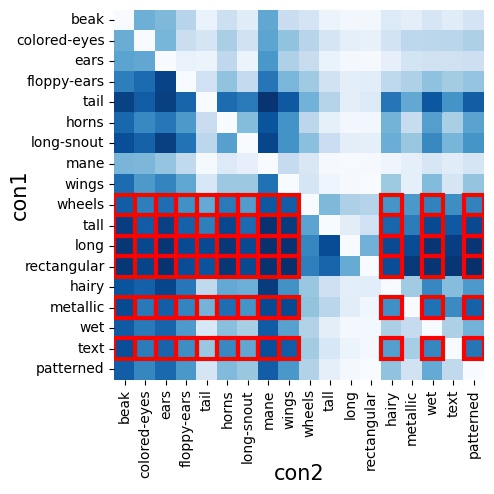}
        \vspace{-0.25in}
        \caption{Ground-truth heatmap for all perturbed images (PGD $l_{\infty}$). }
        \label{sfig:rf_pgd}
    \end{subfigure}
    \hspace{0.05in}
    \begin{subfigure}[t]{0.22\linewidth}
        \centering
        \includegraphics[width=\linewidth]{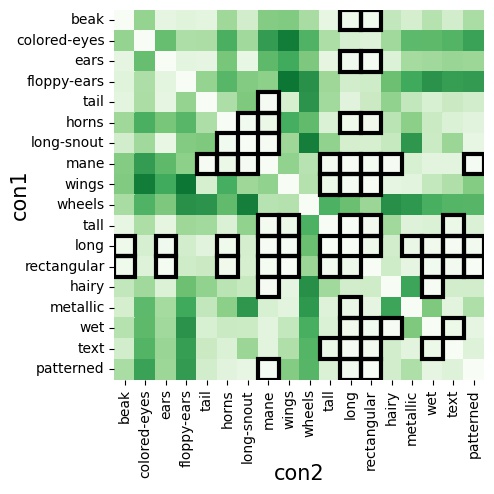}
        \vspace{-0.25in}
        \caption{Differential heatmap for absolute difference between Figures~\ref{sfig:correct_truck} and~\ref{sfig:rf_pgd}.}
        \label{sfig:rf_diff}
    \end{subfigure}
    \hspace{0.05in}
    \begin{subfigure}[t]{0.22\linewidth}
        \centering
        \includegraphics[width=\linewidth]{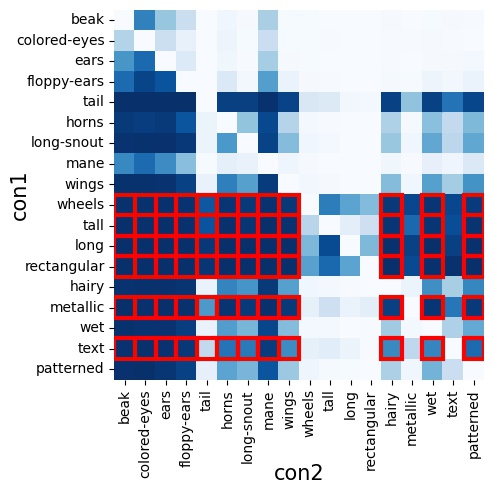}
        \vspace{-0.25in}
        \caption{Ground-truth heatmap for all perturbed images (PGD  $l_2$). }
        \label{sfig:rf_pgd2}
    \end{subfigure}
    \hspace{0.05in}
    \begin{subfigure}[t]{0.22\linewidth}
        \centering
        \includegraphics[width=\linewidth]{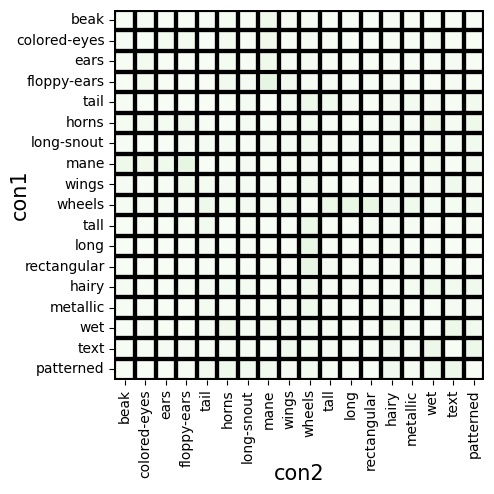}
        \vspace{-0.25in}
        \caption{Differential heatmap for absolute difference between Figures~\ref{sfig:correct_truck} and~\ref{sfig:rf_pgd2}.}
        \label{sfig:rf_diff2}
    \end{subfigure}
   
    \caption{Heatmaps for identifying robust and non-robust features against Projected Gradient Descent (PGD) attack (both $l_{\infty}$ and $l_2$). All heatmaps consider images with ground truth \textit{truck}. The black outline shows robust strength predicates where the absolute difference in satisfaction probability $\leq0.05$. The red outline indicates relevant strength predicates for class \textit{truck}.}
    \label{fig:rf}
\end{figure*}

\begin{figure*}[t]
    \centering
    \begin{subfigure}[t]{0.22\linewidth}
        \centering
        \includegraphics[width=\linewidth]{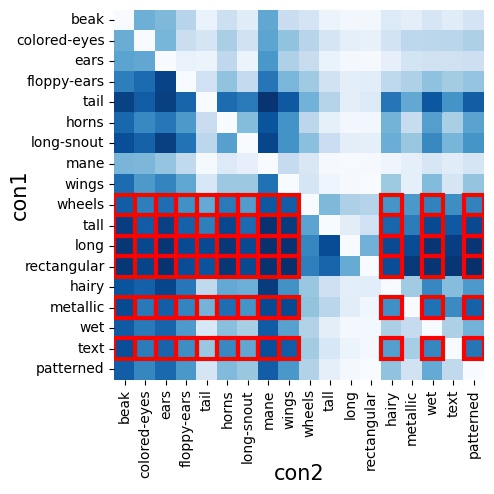}
        \vspace{-0.25in}
        \caption{Ground-truth heatmap for all perturbed images (PGD  $l_{\infty}$). }
        \label{sfig:rf_pgd}
    \end{subfigure}
    \hspace{0.05in}
    \begin{subfigure}[t]{0.22\linewidth}
        \centering
        \includegraphics[width=\linewidth]{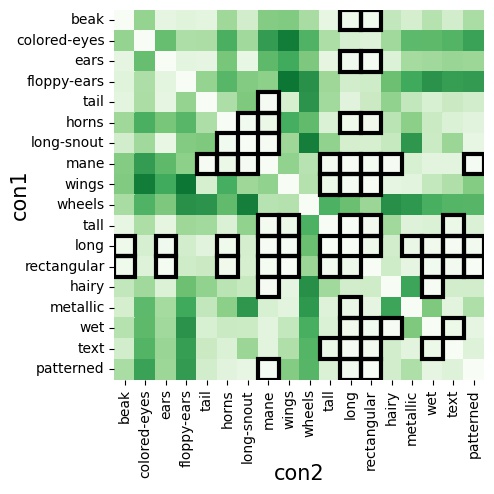}
        \vspace{-0.25in}
        \caption{Differential heatmap for absolute difference between Figures~\ref{sfig:correct_truck} and~\ref{sfig:rf_pgd}.}
        \label{sfig:rf_diff}
    \end{subfigure}
    \hspace{0.05in}
    \begin{subfigure}[t]{0.22\linewidth}
        \centering
        \includegraphics[width=\linewidth]{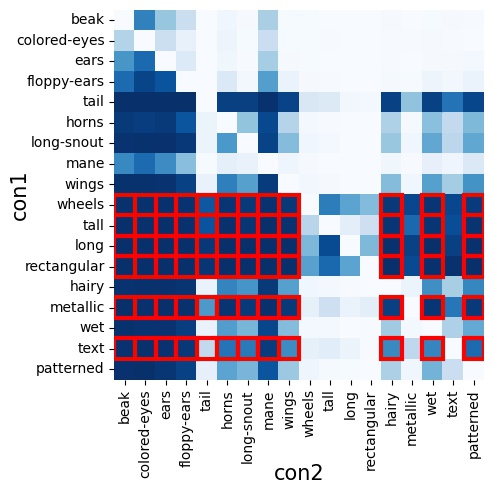}
        \vspace{-0.25in}
        \caption{Ground-truth heatmap for all perturbed images (PGD  $l_2$). }
        \label{sfig:rf_pgd2}
    \end{subfigure}
    \hspace{0.05in}
    \begin{subfigure}[t]{0.22\linewidth}
        \centering
        \includegraphics[width=\linewidth]{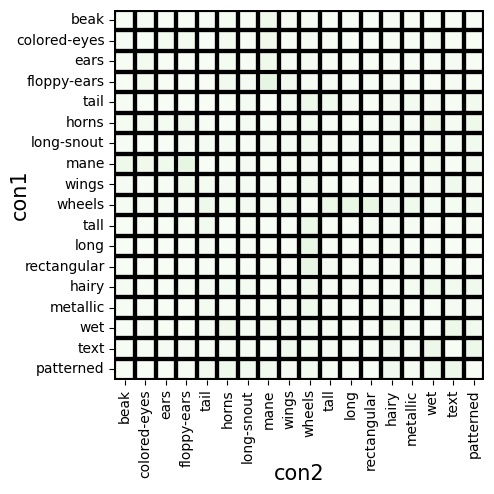}
        \vspace{-0.25in}
        \caption{Differential heatmap for absolute difference between Figures~\ref{sfig:correct_truck} and~\ref{sfig:rf_pgd2}.}
        \label{sfig:rf_diff2}
    \end{subfigure}
    
    \caption{Heatmaps for identifying robust and non-robust features against Projected Gradient Descent (PGD) attack (both $l_{\infty}$ and $l_2$). All heatmaps consider images with ground truth \textit{truck}. The black outline shows robust strength predicates where the absolute difference in satisfaction probability $\leq0.05$. The red outline indicates relevant strength predicates for class \textit{truck}.}
    \label{fig:rf}
\end{figure*}

%\subsection{Application 3: Identification of Robust and Non-robust Strength Predicates}
\subsection{Debugging Vulnerabilities}
%Identification of Robust and Non-robust Strength Predicates}
\label{subsec:robust_feat}

To analyze the model with respect to adversarial examples, we generated adversarial images for every image from the RIVAL10 test set. We perturbed each image with Projected Gradient Descent (PGD) (perturbation radius $ \epsilon=\frac{8}{255}$)~\cite{MadryMSTV18}. We considered both $l_{\infty}$ and $l_{2}$ PGD attacks. 
%We applied our localization algorithm to localize encoder vs head errors for the adversarial inputs; 90\% of all the mis-classifications were categorized as encoder errors.

ResNet18 has $0\%$ accuracy on all the images perturbed with the $l_{\infty}$ PGD, i.e., it is not robust against the $l_{\infty}$ PGD attack. We applied our localization technique to classify the errors as encoder vs head errors; 93.27\% of all the errors were categorized as encoder errors.
Figure~\ref{fig:all_truck}  shows the ground-truth summary heatmap obtained with all original \textit{truck} images without any perturbations, and Figure~\ref{sfig:rf_pgd} shows the heatmap obtained with all \textit{truck} images perturbed with the $l_{\infty}$ PGD attack.
%ResNet18 has a $0\%$ accuracy on all the perturbed images, i.e., it is not robust against the $l_{\infty}$ PGD attack. 
The heatmaps indicate that the satisfaction of most strength predicates is affected by the perturbation. By visualizing the absolute difference between these two heatmaps (see Figure~\ref{sfig:rf_diff}, lighter green indicates a smaller difference), we can identify a small set of predicates that remained constant, i.e., with absolute difference $\leq 0.05$, that are robust predicates, e.g., %$ears\strength beak
{\em rectangular $\strength$ horns, long $\strength$ ears}, and the rest are non-robust predicates, e.g., {\em wheels $\strength$  tail,~%rectangular\strength tail$
metallic $\strength$ long-snout}. Notice that the majority of all relevant predicates for \textit{truck} are not robust, suggesting non-robust features.
%ResNet18 is not robust against $l_{\infty}$ PGD.
%\divya{mention some more robust and non-robust features}. 

Considering the $l_{2}$ PGD attack, ResNet18 is robust---it has a $96.3\%$ accuracy on perturbed images. In addition to high accuracy, we can validate using the heatmap obtained for all \textit{truck} images perturbed with $l_{2}$ PGD attack (Figure~\ref{sfig:rf_pgd2}) as well as the differential heatmap between Figure~\ref{fig:all_truck} and Figure~\ref{sfig:rf_pgd2}
that all relevant strength predicates are satisfied. %that ResNet18 can robustly satisfy all of the relevant strength predicates. 
%In this case, all the non-robust predicates have satisfaction probability $0$ before and after the perturbation. \ravi{i dont understand why these are being called non-robust? Isnt a predicate non-robust only if there is a big change in satisfaction probability after perturbation?} 
%\divya{point out non-robust features if any. Some irrelevant predicates have a large difference example floppy-ears - mane, rectangular - wheels. Mention that these do not impact model output for truck.} 

Analyzing model robustness through concepts, we can assign meaning to statistical results, allowing us to better understand the effects of the adversarial perturbations, and thus allowing developers to design more meaningful detectors and attacks. This experiment also highlights that most adversarial examples get misclassified due to encoder errors. While the encoder of ResNet18 is nearly accurate for in-distribution inputs, adversarial perturbations can lead to incorrect encoding as suggested by the non-robust features. Such insights can facilitate re-training or repair efforts by directing attention to the encoder and the specific non-robust strength predicates.

%Tail is a concept, $tail > metallic$ is a feature. 
%Identify invariant properties. 
%Can help re-training with focus on the non-robust features. 

%Experiment:
%\begin{itemize}
%    \item plot heatmaps w.r.t. ground truth before and after perturbations, 
%    \item highlight features ``preserved'' (diff $\leq threshold$) 
%\end{itemize}

%\blue{Results are shown in Fig.~\ref{fig:rf}.}

%Results we have: (adv) pgd, fab, fgsm, pgdl2, mix, (natural) brightness%, (poison) gradient matching, \red{can run more if needed}

\begin{figure*}[t]
    \centering
    \begin{tabular}{ccc}
    \begin{subfigure}[t]{0.22\linewidth}
        \centering
        \includegraphics[width=\linewidth]{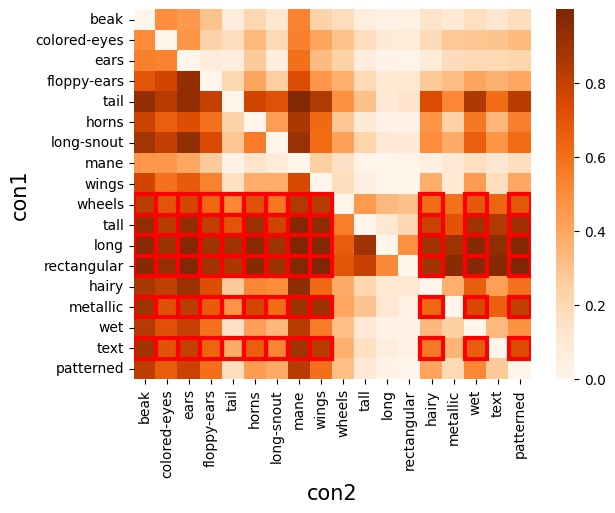}
    \vspace{-0.25in}
    \caption{Output-label summary heatmap for perturbed images classified as \textit{truck}.}
    \label{sfig:detect_pgd}
    \end{subfigure} &
    \begin{subfigure}[t]{0.22\linewidth}
        \centering
        \includegraphics[width=\linewidth]{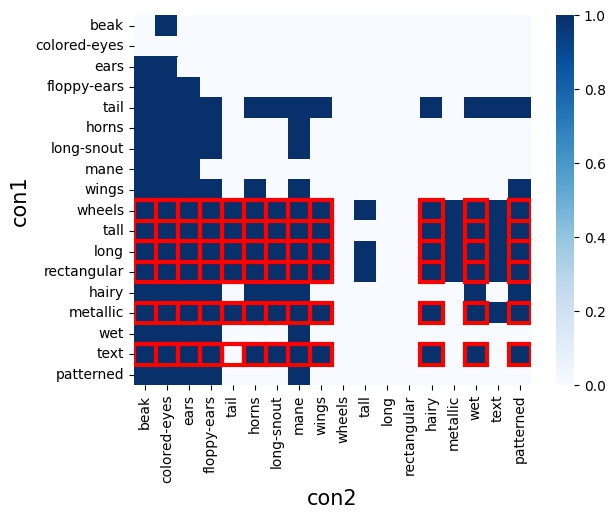}
    \vspace{-0.25in}
    \caption{Binarized heatmap of correctly classified clean images of \textit{trucks} (binarized Figure~\ref{sfig:correct_truck}). }
    \label{sfig:detect_c}
    \end{subfigure}&
    \begin{subfigure}[t]{0.22\linewidth}
        \centering
        \includegraphics[width=\linewidth]{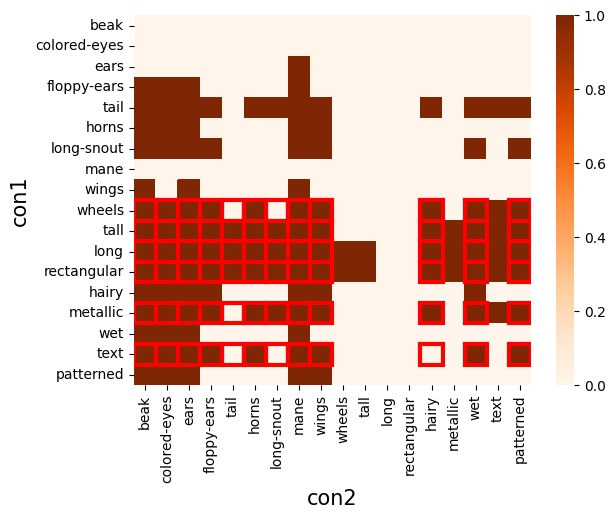}
        \vspace{-0.25in}
        \caption{Binarized heatmap of perturbed images classified as \textit{truck} (binarized Figure~\ref{sfig:detect_pgd}). }
        \label{sfig:detect_pgd_b}
    \end{subfigure}\\
    \end{tabular}
    \caption{Heatmaps used for runtime detection of adversarial inputs. Adversarial inputs generated with $l_{\infty}$ PGD attack, $\epsilon=\frac{8}{255}$. Figure~\ref{sfig:detect_c} and Figure~\ref{sfig:detect_pgd_b} binarized with threshold $0.6$. Red outlines indicate relevant strength predicates for class \textit{truck}. }
    \label{fig:rd}
    %\vspace{-0.1in}
\end{figure*} 

\begin{table*}[t]
    \centering
    \caption{Runtime Defect Detection Accuracy: %Columns PGD-Mixture show the accuracies for detecting adversarial inputs vs clean inputs with heatmaps for all $5000$ images in the test set. 
    %\caroline{
    %We used $75\%$ clean test images and their perturbed versions to obtain heatmaps offline, and the rest $25\%$ clean images and perturbed images are used for evaluation. 
    %For each class $c$, we have about 500 clean images with ground truth $c$ and 500 perturbed images that are labeled by ResNet18 as $c$. 
    First five columns show results for adversarial inputs, each cell contains ($a_p$, $a_n$) where $a_p$ is the percentage of perturbed images correctly detected as adversarial inputs ($IoU_p > IoU_n$), and $a_n$ percentage of clean images correctly detected ($IoU_p \leq IoU_n$). $\epsilon$ is $\frac{8}{255}$ for all experiments. Last column shows results for misclassifications; ($a_m$, $a_c$), where $a_m$ is the \% misclassifications correctly detected and $a_c$ is the \% correct classifications not flagged as faulty.}
    \label{tab:adv_acc}
    \vspace{0.05in}
    \begin{tabular}{cccccc|c}
    \toprule
        & \shortstack{PGD $l_{\infty}$} & \shortstack{FAB $l_{\infty}$} & \shortstack{FGSM $l_{\infty}$} & \shortstack{PGD $l_{2}$} & \shortstack{Mixture} & \shortstack{Misclassifications}\\
         \midrule
         \textit{truck} & (0.519,0.898) & (0.621,0.961) & (0.897,0.916) & (0.234,0.782) & (0.528,0.808) & (0.80 , 0.610)\\
\textit{car} & (0.55,0.964) & (0.541,0.992) & (0.807,0.906) & (0.083,0.926) & (0.583,0.944) & (0.50 , 0.846)\\
\textit{plane} & (0.656,0.884) & (0.595,0.97) & (0.745,0.96) & (0.032,0.973) & (0.735,0.927) & (1.0 , 0.921)\\
\textit{ship} & (0.748,0.891) & (0.724,0.938) & (0.934,0.966) & (0.163,0.853) & (0.794,0.907) & (0.50 , 0.427)\\
\textit{cat} & (0.954,0.447) & (0.76,0.94) & (0.684,0.914) & (0.052,0.956) & (0.756,0.54) & (1.0 , 0.938)\\
\textit{dog} & (0.535,0.788) & (0.634,0.947) & (0.567,0.955) & (0.04,0.965) & (0.46,0.881) & (0.75 , 0.867)\\
\textit{equine} & (0.808,0.624) & (0.815,0.917) & (0.858,0.925) & (0.159,0.887) & (0.705,0.827) & (0.50 , 0.077)\\
\textit{deer} & (0.58,0.922) & (0.714,0.94) & (0.581,0.962) & (0.048,0.967) & (0.713,0.909) & (0.667 , 0.855)\\
\textit{frog} & (0.721,0.798) & (0.595,0.932) & (0.692,0.965) & (0.123,0.886) & (0.38,0.957) & (0.667 , 0.718)\\
\textit{bird} & (0.679,0.71) & (0.794,0.968) & (0.554,0.965) & (0.033,0.968) & (0.611,0.767) & (1.0 , 0.961)\\
\hline
Total & (0.678,0.792) & (0.683,0.95) & (0.74,0.943) & (0.097,0.917) & (0.631,0.847) & (0.163 , 0.951)\\
         %\bottomrule
         %\multicolumn{5}{l}{*All numbers are rounded.}
    \end{tabular}
    \vspace{-0.1in}
\end{table*}

\subsection{Runtime Defect Detection}
\label{subsec:adv_detect}
%\caroline{Summary of this paragraph: run time detection can be done by comparing each single-input heatmap with summary heatmaps of 1) correctly classified original images of a class; 2) perturbed images misclassified to this class. The comparison can be done by simply binarizing summary heatmaps with a threshold, and using IoU to determine whether each single-input heatmap is more similar to heatmap 1) or heatmap 2). Using CLIP directly can give more accurate results, but CLIP is too expensive for runtime detection. }

%\corina{do we explain anywhere what adversarial inputs are and why it is important to detect them??}

So far, we have primarily relied on the ground-truth summary heatmaps for interpreting model behaviour. When ground truth information is unavailable, e.g., during runtime, we can utilize output-label summary heatmaps for analysis, enabling runtime detection of defects such as adversarial inputs and misclassifications. 

\subsubsection{Runtime Detection of Adversarial Inputs} 
%[maybe cut this paragraph]
As we observed before, adversarial perturbations lead to encoder errors. Therefore, with output-label summary heatmaps of a particular class, the relevant predicates would appear satisfied. 
%Intuitively, perturbations can affect the encoding of concepts in the inputs, therefore, heatmaps for non-perturbed images of one class would be different from heatmaps for perturbed images classified to the same class.
%Adv detectionL since adv perturbations lead to encoder errors, when look at output based heatmaps for a label, the relevant predicate gets satisfied by both miusclassifie and correctly classified; but the not-relevant are different. Since they were mapped from other classes, giving random satisfaction rate. We zoom in on this, thus we use single-output and then …. Reason this relevant seems same with perturbat, previously, adv lead to lots of encoder errors.
For example, consider the PGD $l_{\infty}$  attack ($\epsilon=\frac{8}{255}$); Figure~\ref{sfig:detect_pgd} shows the summary heatmap for all perturbed images misclassified to \textit{truck} by ResNet18. Most relevant predicates for \textit{truck} are satisfied. %since these images are classified as \textit{trucks}. 
However, since these images are in fact incorrectly mapped to \textit{trucks} (from different ground-truth classes), the predicates that are not relevant to \textit{truck} but that are relevant to other classes, show a more different and random satisfaction rate. Zooming in on this for adversarial detection, for each input, we identify such differences in its single-input heatmap.%\divya{single-input heatmap} 
%Compared to Fig.~\ref{sfig:rf_c}, the satisfaction of most strength predicates is affected by the perturbation and differs from that of non-perturbed images. 
%We demonstrate that adversarial inputs can be detected during runtime by identifying such differences.

To this end, we leverage a held-out dataset %\corina{what is this exactly?? train set??} 
to construct, binarized ground-truth summary heatmaps and output summary heatmaps for clean and perturbed images classified to the same class by the model. Then, for each input encountered at runtime with output class $c$, it is considered adversarial if its single-input heatmap is more similar to the perturbed heatmap than the clean heatmap of class $c$, and clean otherwise. The similarity is computed using Intersection over Union (IoU).

%To determine the similarity between a single-input heatmap and the above summary heatmaps,
%\susmit{is summary over all of training data - different inputs have different heatmaps}\caroline{addressed}, 
%we first binarize the summary heatmap with a threshold $t$ 
%\susmit{some insight into selection of t would be useful}\caroline{added after this sentence, please check} 
%to highlight strength predicates with high satisfaction probabilities, i.e., $1$ if satisfaction probability $\geq t$, and $0$ otherwise. 
%The value of $t$ should be sufficiently high to capture the primary differences between clean and perturbed heatmaps while avoiding the inclusion of small values that are predicates satisfied by only a small set of images. 
%In our experiment, we set 
%$t=0.6$, which resulted in the highest detection accuracy.

We constructed sets of perturbed inputs ($\epsilon=\frac{8}{255}$) using common adversarial attacks such as 
PGD (both $l_{\infty}$ and $l_2$)~\cite{MadryMSTV18}, FGSM ($l_{\infty}$)~\cite{Szegedy2014IntriguingPO}, FAB ($l_{\infty}$)~\cite{croce2020minimally}, and a random mixture of all the above. One perturbed image was created for every input in the RIVAL10 test set. 
We used $75\%$ clean test images and their perturbed versions to obtain the heatmaps offline, and the rest $25\%$ clean images and perturbed images are used for runtime evaluation. 
%We used a held-out dataset with clean and perturbed inputs to construct the binarized heatmaps for clean and peturbed respectively. We selected a threshold of $t=0.6$.
%We chose a threshold $t>0.5$ for binarization ($t=0.6$ in our experiments). \corina{shall we add more? say that other thresholds lead to similar results??}
The threshold $t$ should be sufficiently high to capture the differences between clean and perturbed heatmaps while avoiding the inclusion of predicates satisfied by only a small set. In our experiment, we set $t=0.6$, which resulted in the highest detection accuracy.
The binarized clean heatmap and the binarized perturbed heatmap for class \textit{truck} are shown in Figure~\ref{sfig:detect_c} and Figure~\ref{sfig:detect_pgd_b} respectively. 

%The binarized clean heatmap (binarized Figure~\ref{sfig:correct_truck}) is shown in Figure~\ref{sfig:detect_c}; and the binarized perturbed heatmap (binarized Figure~\ref{sfig:detect_pgd}) is shown in Figure~\ref{sfig:detect_pgd_b}, with threshold $t=0.6$. 
%Then we compute Intersection over Union (IoU) as the similarity metric, between each single-input heatmap and the binarized summary heatmaps; we use $IoU_p$ for similarity with the the binarized perturbed heatmap, and $IoU_n$ for similarity with the binarized clean heatmap. An input with $IoU_p > IoU_n$ suggests it is adversarial, and clean otherwise.

%To demonstrate runtime detection of adversarial inputs, 
%%we considered common adversarial perturbations 
%we constructed sets of perturbed inputs (all with $\epsilon=\frac{8}{255}$) using common adversarial attacks such as 
%PGD (both $l_{\infty}$ and $l_2$)~\cite{MadryMSTV18}, FGSM ($l_{\infty}$)~\cite{Szegedy2014IntriguingPO}, FAB ($l_{\infty}$)~\cite{croce2020minimally}, and a random mixture of all the above. 
%\corina{I assume these are all perturbations of the test set??}

In Tbl.~\ref{tab:adv_acc}, in the first five columns we show the accuracy of identifying perturbed and clean images per class. Overall, we achieved  high accuracy for most classes and most perturbations, showing the potential of our lightweight approach.
%. This suggests our heatmaps have the potential for more precise runtime detection. 
The detection accuracy for PGD $l_2$ is low. As we discussed in Sec.~\ref{subsec:robust_feat}, ResNet18 is highly robust to PGD $l_2$ attacks, thus the perturbed and clean heatmaps are too similar to enable differentiating adversarial inputs. 

%\corina{not sure this is interesting??}Additionally, we noticed that adversarial attacks affect classes differently---classes with lower accuracy have more similar perturbed and clean heatmaps, suggesting the model is more robust for some classes of inputs than others. 
%Overall these results suggest that our adversarial detector works better when the model is less robust since the clean and adversarial heatmaps are very similar in these cases.

%we only used images clip correctly classified. Attack works differently for different classes: classes with low accuracy have similar heatmaps between original and perturbed data (model is robust).

\subsubsection{Runtime Detection of Misclassified Inputs}
%[corina: not sure this is the right terminology]
The same method as outlined above can also be used to monitor for inputs that are not perturbed but may still be misclassified, due to model inaccuracies. The challenge is that the goundtruth is not known at runtime; we can leverage semantic heatmaps as follows. We use an offline dataset comprising of in-distribution data that are classified correctly and mis-classified by the ResNet18 model to construct the respective binarized output-label summary heatmaps. At runtime, the single input heatmap of every new input is compared with the correct and misclassified heatmaps to detect mis-classifications. The last column of Table~\ref{tab:adv_acc} shows the results. We constructed the offline dataset with 75\% of correctly-classified inputs and 75\% of the mis-classified inputs from the RIVAL10 test dataset. A dataset comprising of the remaining 25\% of correctly classified and misclassified inputs were employed at runtime to obtain the accuracy results. Note that since the in-distribution accuracy of ResNet18 is high, there were very few misclassifications compared to correct classifications in the offline dataset. This impacts the precision of the heatmaps, which in turn impacts the total detection accuracy. However, the misclassification detection accuracies for specific labels are high (such as 80\% for truck and 75\% for dog).

Note that VLMs such as CLIP can themselves be deployed as runtime detectors to catch inputs misclassified by the vision model. %CLIP has robust accuracy $0.982$ against PGD $l_{\infty}$, $0.984$ against FAB, $0.973$ against FGSM, and $0.984$ against PGD $l_{2}$, all of which are higher than detection accuracy with semantic heatmaps. 
However, foundation models such as CLIP are too computationally heavy to be deployed during runtime, while our heatmaps can be precomputed, only the single-layer linear aligner $r_{map}$ is required during runtime, which enables a more lightweight analysis. %which is more desirable for analysis during runtime. 

\section{Threats to Validity}
\label{sec:threats}
%\todo{}
%- we only looked at one model; why? why do/don't we think our findings to generalize to other models and datasets? \\
%- we only used one VLM?

%\corina{added this not sure}

Our approach relies on the set of human-defined concepts which may be incorrect. In this paper, we use high-quality information about concepts from a previous study~\cite{moayeri2022comprehensive}. 
%Our approach relies on concept annotations provided by humans, and incorrect annotations can affect the analysis. In this paper, we use the existing dataset RIVAL10~\cite{moayeri2022comprehensive} with concept annotations of high quality~\cite{mangal2024conceptbased}. 
%\caroline{reworded a bit}. 
%\corina{not sure} 
In the future, we plan to explore VLMs to {\em elicit} such concepts automatically. %Note that we do not require manual annotations for each image.
Note that our approach does not require images annotated with concepts.

We only demonstrated our approach on one (albeit complex) model and one dataset.  Thus we can not claim that the results would generalize to other models and/or other datasets. More work is planned for the future that involves other image datasets and vision models. However, we note that the methodology and the analysis applications presented in this paper are generic and are not tied to a particular model or DNN architecture.

%\corina{moved this bit from conclusions}

The choice of VLM is vital to our approach, as it limits us to inputs the VLM correctly classifies and perturbations to which the VLM is robust. During our case study, we noticed that the version of CLIP used in our experiments is not robust to natural perturbations such as brightness (with a robust accuracy of $14.8\%$). Additionally, our approach depends on the quality of $r_{map}$. To achieve an effective mapping, the dimension of the representation space of the vision model should be similar to that of the VLM. 
%We leave exploring other VLMs for natural perturbations and other mapping techniques between the representation spaces for future work. 
We leave exploring other VLMs to build semantic maps and developing other mapping techniques between the representation spaces for future work. 
\section{Related Work}
\label{sec:related_work}

% \begin{itemize}
%     \item concepts [the previous paper]
%     \item robust vs non-robust features~\cite{kalimeris2019sgd,ilyas2019adversarial,zhou2021towards}
%     \item debugging~\cite{zhang2022drml,eyuboglu2022domino,JainLMM23}
%     \item fault localization~\cite{GhanbariTAR23,WardatLR21}
%     \item ...
% \end{itemize}

\noindent{\textit{\textbf{Concept-based  Analysis of Neural Networks.}}}
Closest to our work are the ideas presented in \cite{mangal2024conceptbased} including the $\conspec$ specification language with the notion of concept-based strength predicates. While~\cite{mangal2024conceptbased} aims for formal verification of properties expressed $\conspec$, we aim for statistical analysis of concept predicates, which in turn enables novel techniques for  fault localization and runtime monitoring.
The field of concept representation analysis in neural networks, as explored in \cite{Zhou_2018_ECCV, GhorbaniWZK19, pmlr-v80-kim18d, DBLP:series/faia/YehKR21,crabbe2022concept}, seeks to extract representations of concepts in the representation spaces of neural networks, and use these extracted representations to quantify the impact of high-level concepts on model output via attribution-based techniques. However these techniques usually require manual annotations for each image in a test set.
There is also a growing body of work on extracting concept representations learned by language models in order to understand and control them~\cite{bai2023concept, nanda2023emergent,wang2023concept,park2023linear,huben2024sparse,zou2023representation}. %\caroline{I feel like how we are different is missing here. Would these be appropriate: concepts overlap and are more quantitative than boolean (from intro); also our analysis is based on visualization of the concepts.}\corina{i tried to address}

%\vspace{0.05in}
\noindent\textit{\textbf{VLMs for Analyzing Vision Models.}}
%Existing work ~\cite{zhang2022drml,eyuboglu2022domino,JainLMM23}, highlight the potential of using VLMs for testing and debugging ; but either use the VLM as encoder, or perform black-box test gen for other vision models.
%Recent work~\cite{moayeri2023text} demonstrates that VLMs can help probe other vision models more intricately. highlight the use-cases presented in this work. 
Recent work~\cite{moayeri2023text} first observed that the representation spaces of vision models and VLMs can be linked to each other via a linear map. They used this mapping for a number of use-cases---extending the vision model to identify new classes in a zero-shot fashion,  diagnosing data distribution shifts in
terms of human concepts, retrieving images
satisfying a set of text-based constraints, and so on. However, they did not propose to leverage this mapping for the analysis of vision models as we do here. Other recent works~\cite{zhang2022drml,eyuboglu2022domino,JainLMM23} have proposed the use of VLMs to find a coherent, human-understandable description for inputs that are misclassified by a vision model. Given such a set of misclassified inputs, these techniques directly use the VLM to extract coherent textual descriptions that characterize such inputs. Unlike our semantic maps, these techniques do not seek to understand the internal behavior of vision models in a human-understandable fashion.

%\vspace{0.05in}
%\noindent\textit{\textbf{Explaining Neural Network Behavior.}}
%Work on explaining neural networks can be broadly divided between techniques that aim to find the input features with the largest effect on model behavior such as gradient-based~\cite{simonyan2014deep, sundararajan2017axiomatic} and activation-based~\cite{46832, Gradcam} attributions; and those that aim to explain the internal reasoning performed by a model (see~\cite{rauker2023toward,milliere2024philosophical} for excellent surveys on the latter). Techniques that focus on internal reasoning seek to explain the representations learned and the computations performed by a model in a human-understandable form. 
%Our work on using semantic heatmaps for explaining model behavior falls under this category.

%\vspace{0.05in}
\noindent\textit{\textbf{Debugging for Neural Networks.}}
Techniques for fault localization have been proposed to localize errors in trained models \cite{ma2018mode,eniser2019deepfault,GhanbariTAR23} or in programs and libraries that are used to build the models \cite{DeepLocalize,DeepDiagnosis,
NeuraLint}. Our work is closest to the former. One closely related work \cite{FahmyPBB21} also uses heatmaps  to group inputs that have similar neuron profiles at inner layers.  These heatmaps  help identify portions of input or neurons that impact the output, but rely on visual interpretability to understand the results. In contrast, we use heatmaps in terms of  concepts in the VLM embedding space. 
We see our work as complementary to other techniques. Our debugging and semantic heatmaps can help give a global view of the reason for failure, in terms of human-understandable concepts.
%OLD:
%using spectrum\hskip0pt-\hskip0pt based fault localization that can localize faults in neural networks to specific neurons~\cite{ma2018mode,eniser2019deepfault} and those based on mutation-based fault localization that can localize faults to specific layers~\cite{GhanbariTAR23} have been proposed in the literature. These techniques statistically analyse model behavior on sets of passing and failing inputs in order to isolate the faulty component. 
%\corina{this seems wrong here as previous techniques can do the same}
%In contrast, our approach is able to isolate the faulty component even for individual failing inputs. Moreover, unlike these past approaches, our semantic heatmaps can help understand the reason for faulty behavior in human-understandable terms. 
Recently, model editing~\cite{meng2022locating,meng2023massediting} techniques have been proposed that perform causal analysis, by intervening on neural activations, to isolate neurons in language models that lead to specific buggy behaviors. 
%The behavior of such isolated neurons is semantically interpretable and therefore, these techniques have the potential to perform human-understandable fault localization in the context of generative language models. 
These techniques can be expensive in practice and we view them as complementary to our approach.
\noindent\textit{\textbf{Runtime Detection.}}
Previous techniques for runtime detection of misclassifications, see e.g.,~\cite{ConfidNet,SelfChecker,DeepInfer,MisPred}, %\corina{to add citations [tbd]})
are generally based on the idea of capturing a training profile (probability distributions, confidence values, or data preconditions) and measure differences for new inputs. In our case we leverage heatmaps to capture correct/incorrect execution profiles and measure semantic difference for new inputs. 
 There is a large body of literature on detecting adversarial examples (see~\cite{aldahdooh2022adversarial} for a survey). We take a more semantic approach, that discovers and leverages robust and non-robust features for detection. 
 
%A number of approaches have been proposed to evaluate the reliability of a DNNs in deployment~\cite{ConfidNet,SelfChecker,DeepInfer,MisPred}. SelfChecker~\cite{SelfChecker} computes the similarity between DNN layer features of unseen instances and the samples in the training set, using kernel density estimation
%(KDE) to detect misclassifications by the model in deployment.
%\corina{rewrote: please check; all the citations shoudl be discussed??} DeepInfer~\cite{DeepInfer} describes a white-box analysis that computes and propagates preconditions over a DNN's  layers and uses them to determine the model’s correct or incorrect prediction. Existing approached are limited to tabular data (DeepInfer) or simple image classification tasks (SelfChecker).
%In contrast, the approach we present here aims to identify failure-inducing image inputs based on an analysis of high-level concepts. It is more lightweight and more widely applicable than existing methods, that use computationally-intensive operations over the internals of the models.

\section{Conclusion and Future Work}
\label{sec:conclusion}

In this paper, we proposed novel techniques for debugging and runtime defect detection for vision models. The techniques leverage a separate VLM and  \emph{semantic heatmaps} that summarize statistical properties of DNNs using human-understandable concepts. 
We demonstrated the proposed techniques through a case study with a ResNet classifier and CLIP.
%, we demonstrated the potential of our heatmaps for semantic analysis of DNN behaviors, including explaining model behavior, debugging and analysis of robust features during design-time (offline), and detecting adversarial inputs during run-time. 
%Our approach helps visualize model behavior, thus aids in the development of effective and interpretable analysis of DNNs, and contributes to the longer-term goal of safety assurance of DNNs and DNN-enabled systems.
%\ravi{Should we move this to conclusion?}\caroline{moved}
%\susmit{behavior or reasoning  - we could use a common term throughout the paper}\caroline{Thank you for the suggestion. I will do a pass to unify the terminologies once everyone finishes their pass}

%Each of the presented applications suggests future research directions. First, we plan to investigate the significance of non-relevant strength predicates on model output to better explain model behavior. Next, 
In future work, we aim to develop methods that enable more precise and accurate error localization to specific layers, possibly by learning more complex alignment maps between the model under analysis and the VLM. We also plan to examine the role of robust features in creating more effective adversarial attacks and explore new methods to enhance adversarial detection accuracy. Lastly, we plan to explore how to use our results for localized repair and also for testing of vision models, including test generation and data selection for retraining with violated strength predicates.

%This work was partially supported by DARPA under 
%Contract No. FA8750-23-C-0519, %% Anirban
%and 
%the U.S. Army Research Laboratory Cooperative Research
%Agreement W911NF-17-2-0196. %% Susmit
%Any opinions, findings, conclusions, or recommendations expressed in this paper are those of the authors and do not necessarily reflect the Department of Defense or the United States Government. 

\section*{Acknowledgements}
This work was partly supported by the US Air Force and DARPA under Contract No. FA8750-23-C-0519, and US ARL Cooperative Agreement W911NF-17-2-0196. Any opinions, findings, conclusions, or recommendations expressed in this paper are those of the authors and do not necessarily reflect the Department of Defense, DARPA, or the US Government.

\bibliographystyle{IEEEtran}
%\bibliography{all}
\bibliography{refs}
\end{document}